\documentclass[pre,showpacs,showkeys,preprintnumbers,amsmath,amssymb,superscriptaddress,twocolumn]{revtex4}
\usepackage{graphicx}
\usepackage{amssymb,amsfonts,amsmath}
\usepackage{epstopdf}
\usepackage{color}
\usepackage{bm}
\usepackage[colorlinks=true,linkcolor=blue,citecolor=red]{hyperref}%

\def\X{\bm{X}}
\def\x{\bm{x}}

\def\C{{\mathbb C}}
\def\E{{\mathbb E}}
\def\P{{\mathbb P}}
\def\R{{\mathbb R}}
\def\L{{\mathcal L}}
\def\T{{\mathcal T}}

\def\pa{{\partial\Omega}}

\def\erf{\mathrm{erf}}

\begin{document}

\title{Encounter-based approach to diffusion with resetting}

\author{Ziyad Benkhadaj}
\affiliation{ENS Paris-Saclay, 91190 Gif-sur-Yvette, France}

\author{Denis~S.~Grebenkov}
\email{denis.grebenkov@polytechnique.edu}

\affiliation{Laboratoire de Physique de la Mati\`{e}re Condens\'{e}e (UMR 7643), \\ 
CNRS -- Ecole Polytechnique, IP Paris, 91128 Palaiseau, France}

\date{Received: \today / Revised version: }

\begin{abstract}
An encounter-based approach consists in using the boundary local time
as a proxy for the number of encounters between a diffusing particle
and a target to implement various surface reaction mechanisms on that
target.  In this paper, we investigate the effects of stochastic
resetting onto diffusion-controlled reactions in bounded confining
domains.  We first discuss the effect of position resetting onto the
propagator and related quantities; in this way, we retrieve a number
of earlier results but also provide complementary insights onto them.
Second, we introduce boundary local time resetting and investigate its
impact.  Curiously, we find that this type of resetting does not alter
the conventional propagator governing the diffusive dynamics in the
presence of a partially reactive target with a constant reactivity.
In turn, the generalized propagator for other surface reaction
mechanisms can be significantly affected.  Our general results are
illustrated for diffusion on an interval with reactive endpoints.
Further perspectives and some open problems are discussed.
\end{abstract}



\maketitle

\section{Introduction}

Diffusion-controlled reactions and the related first-passage phenomena
are ubiquitous in nature and industrial applications
\cite{Rice,Lauffenburger,Redner,Schuss,Metzler,Oshanin}.  In a typical
setting, a particle (e.g., a ligand) diffuses towards a specific
location in a confining environment (e.g., a cytoplasm) and attempts
to bind to or react on that target (e.g., a receptor).  The
macroscopic concentration of particles or, equivalently, the survival
probability of a single particle, satisfies a Fokker-Planck equation
with appropriate boundary conditions.  Since the seminal paper by von
Smoluchowski \cite{Smoluchowski17}, diffusion-controlled reactions
have been thoroughly investigated to reveal the respective roles of
the structural complexity of the environment, of the diffusive
dynamics in the bulk, of the location, shape, size and reactivity of
the target, etc.
\cite{Berg77,Weiss86,Condamin07,Grebenkov07,Benichou08,Benichou10,Benichou10b,Benichou11,Bressloff13,Bray13,Benichou14,Godec16,Godec16b,Grebenkov16,Chechkin17,Lanoiselee18,Levernier19}.

Evans and Majumdar have introduced a new aspect into this field --
stochastic resetting, according to which the particle can be
spontaneously relocated into its initial position to re-start
diffusion towards the target \cite{Evans11}.  Such resetting steps
allow the particle to abandon its original random path that could be
too long or even never leading to the target.  For instance, if the
particle diffuses on the positive half-line towards the origin, the
mean first-passage time (FPT) to that target is infinite due to
contributions of too long paths.  In turn, resetting prohibits long
paths and renders the mean FPT finite.  This basic example reveals
that resetting can be beneficial for diffusive search and lead to a
variety of optimization problems.  We emphasize that this resetting
mechanism is independent of the diffusive dynamics and is thus
different from the instantaneous return process \cite{Feller54} and
its extensions (see
\cite{Sherman58,Grigorescu02,Leung08,Ben-Ari09,DeBruyne20,DeBruyne21}
and references therein), in which the process is reset to a random
bulk point after hitting the boundary or crossing a given threshold.

Since its introduction in 2011, various effects of stochastic
resetting onto diffusion-controlled reactions and related
first-passage times have been studied.  For instance, Evans {\it et
al.} extended the above basic setting to deal with a space-dependent
resetting rate, resetting to a random position drawn from a resetting
distribution, a spatial distribution for the absorbing target
\cite{Evans11b}, partial reactivity of the target, the effect of
multiple searchers \cite{Whitehouse13}, and L\'evy flights
\cite{Kusmierz14} (see also
\cite{Mendez21}).  Pal {\it et al.} focused on time-dependent
resetting rate and determined the survival probability under resetting
and optimal resetting rate function \cite{Pal16}.  Reuveni showed that
the relative standard deviation associated with the FPT of an
optimally restarted process (with a constant rate) was always equal to
$1$, independently of the dynamics \cite{Reuveni16}.  A relation to
Michaelis-Menten reaction scheme \cite{Reuveni14} was also discussed.
Pal and Reuveni proposed an elegent general approach to analyze the
effect of resetting with an arbitrarily distributed resetting time
$\delta$ onto the statistics of any first-passage time $\T$
\cite{Pal17}.  In particular, they deduced a simple formula for the
mean value of the FPT $\T_\phi$ under resetting:
\begin{equation}  \label{eq:T_Pal}
\E\{ \T_{\phi} \} = \frac{\E\{ \min\{\T,\delta\}\}}{\P\{ \T < \delta\}} \,,
\end{equation}
where $\E\{\cdot\}$ denotes the expectation.  This approach was
further elaborated by Chechkin and Sokolov \cite{Chechkin18} into a
general renewal scheme that we will employ in this paper.  They
derived another representation for $\E\{ \T_{\phi} \}$ to investigate
the search optimality under resetting (see Eq. (\ref{eq:Tphi_general})
and the related discussion below).  The effect of refractory period on
stochastic resetting was invesigated in \cite{Evans19}.  In the case
of continuous-time random walks with power-law distributed waiting
times between jumps, long-range memory effects can be considerably
altered by resetting \cite{Bodrova20}, leading to peculiar behaviors
of the propagator and the mean-squared displacement (MSD) of the
particle (see also \cite{Maso19} for other insights onto the MSD).
Dahlenburg {\it et al.} introduced random-amplitude stochastic
resetting, in which the diffusing particle may be only partially reset
towards the origin or even overshoot the origin in a resetting step
\cite{Dahlenburg21}.  The role of a bias due to a potential onto the
search optimality under resetting was analyzed \cite{Pal15,Ahmad22}.
In the case of Poissonian resetting, one can go beyond the propagator
and the first-passage time distribution and investigate additive
functionals of a stochastic process, e.g., its residence time.
Meylahn {\it et al.} considered Markov processes with resetting and
derived the rate function of additive functionals characterizing the
likelihood of their fluctuations in the long-time limit
\cite{Meylahn15} (see also \cite{denHollander19,Smith22}).
Finally, an experimental realization of colloidal particle diffusion
with resetting via holographic optical tweezers was reported
\cite{Tal-Friedman20}.  This work allowed to measure the energetic
cost of resetting and to reveal the need for some improvements in
theoretical analysis to account for fundamental constraints on
realistic resetting protocols.  These and many other aspects of
stochastic resetting have been recently reviewed \cite{Evans20}.  Even
though the renewal approach is valid for rather general diffusive
processes, most former works focused on one-dimensional diffusion on a
line or a half-line, while extensions to higher dimensions concerned
the whole space $\R^d$.  In particular, the specific effects related
to restricted diffusion in bounded domains have not been explored yet.

In this paper, we propose to look at the role of stochastic resetting
in the so-called encounter-based approach to diffusion-mediated
surface phenomena \cite{Grebenkov20}.  This approach is based on the
concept of the {\it boundary local time} $\ell_t$, which quantifies
the number of encounters between the diffusing particle and the
boundary up to time $t$.  The diffusive dynamics is entirely
characterized by the full propagator $P(\x,\ell,t|\x_0)$ -- the joint
probability density of the particle position $\X_t$ and its boundary
local time $\ell_t$ at time $t$, given that the particle started from
a point $\x_0$ at time $0$.  Once the full propagator is determined
for a passive (non-reactive) boundary, different surface reaction
mechanisms can be implemented (see Sec. \ref{sec:encounter}).  In this
way, one can retrieve the conventional constant reactivity described
by the Robin boundary condition as a specific model, one among many
others.  Several extensions and applications of the encounter-based
approach have been recently discussed
\cite{Grebenkov20,Grebenkov20b,Grebenkov20c,Grebenkov21a,Grebenkov22,Bressloff22a,Bressloff22b,Grebenkov22f,Bressloff22c}.
Here, we aim at investigating the role of resetting within this
paradigm
\footnote{
At the submission, we discovered a recently published paper
\cite{Bressloff22c}, which undertakes a similar study in the case of
Poissonian resetting.  Even though all our results were obtained
independently, we systematically outline eventual overlaps with
Ref. \cite{Bressloff22c}; in addition, a comparison between two
approaches is given in Sec. \ref{sec:discussion}.}.

\begin{figure}[t!]
\begin{center}
\includegraphics[width=0.49\textwidth]{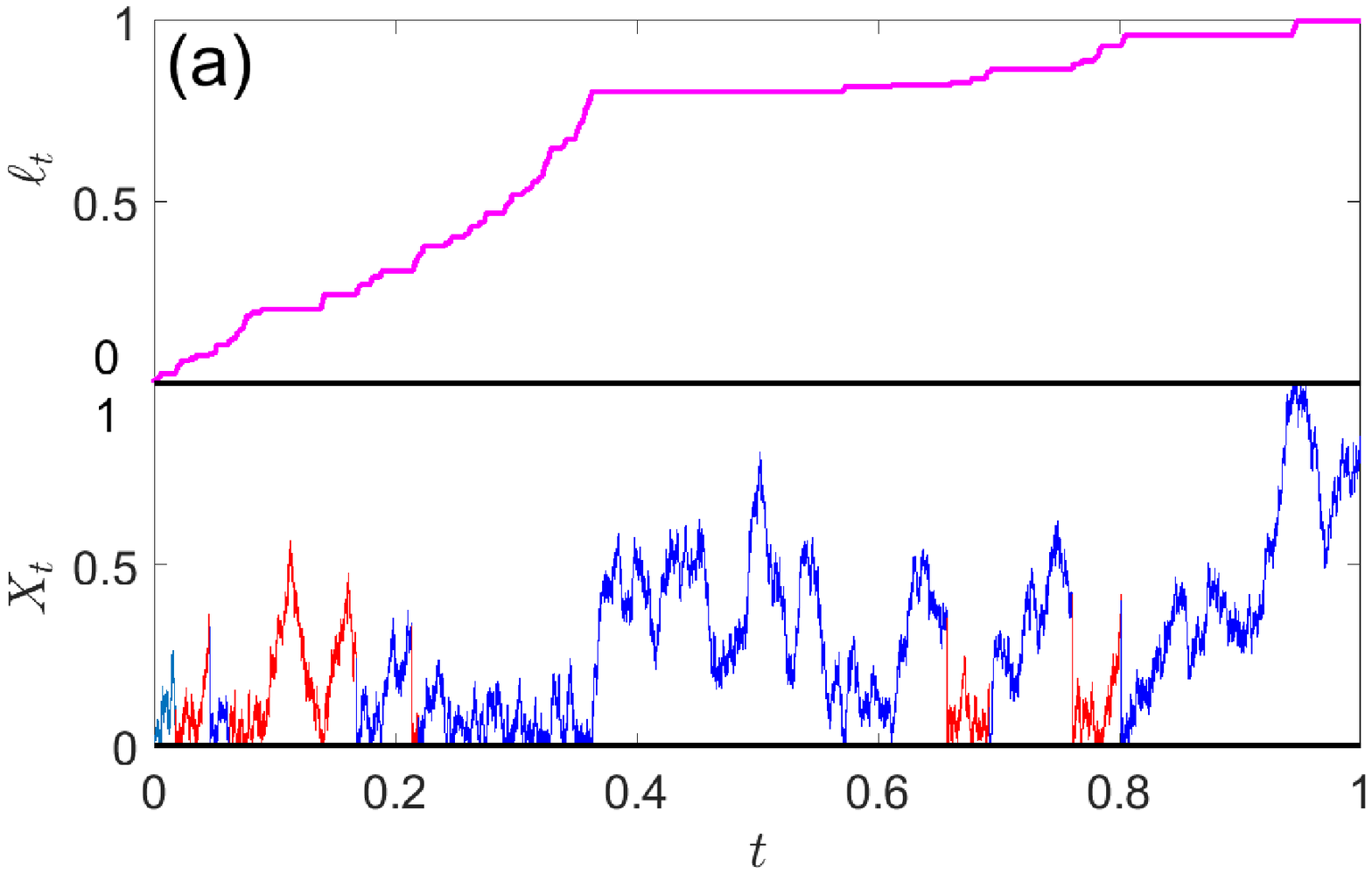} 
\includegraphics[width=0.49\textwidth]{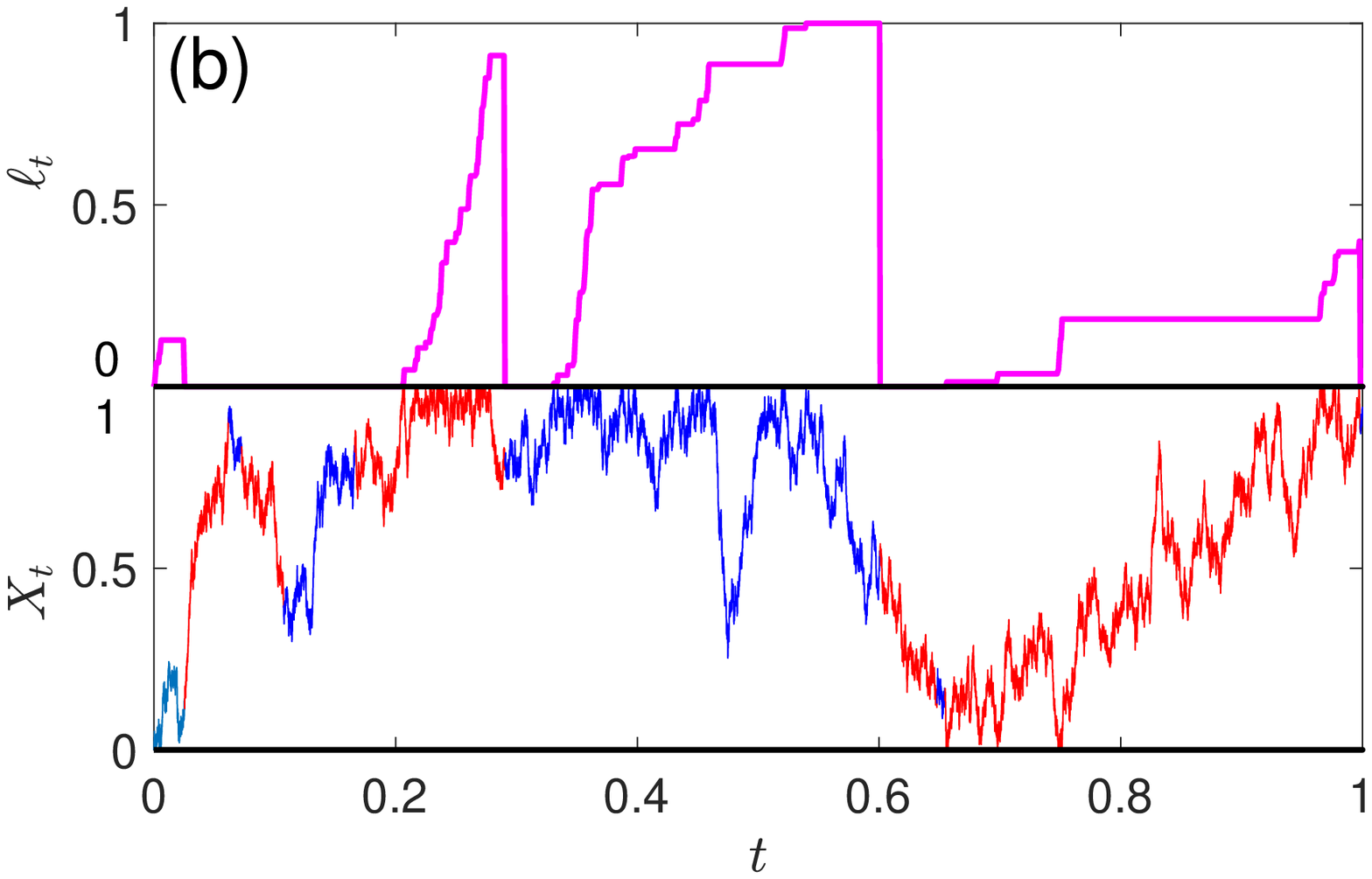} 
\end{center}
\caption{
Simulation of a random trajectory $X_t$ on the unit interval $(0,1)$
and its boundary local time $\ell_t$ under Poissonian resetting with
the law $\P\{\delta > t\} = \Phi(t) = e^{-\omega t}$ for the resetting
time $\delta$, with the rate $\omega = 10$, the diffusion coefficient
$D = 1$, and the starting point $x_0 = 0$.  The boundary local time is
rescaled by its maximum and shifted upwards for an easier
visualization.  {\bf (a)} Position resetting; {\bf (b)} Boundary local
time resetting.  The trajectory is shown by pieces that change color
(blue and red) at each resetting. }
\label{fig:simu}
\end{figure}

Importantly, the encounter-based approach offers more flexibility on
the implementation of resetting (Fig. \ref{fig:simu}).  On one hand,
one can keep the boundary local time as a history of encounters with
the boundary while resetting the position of the particle, as done in
former studies.  In this direction, one should be able to retrieve
former results but also discover new ones, e.g., the distribution of
the boundary local time or correlations between $\X_t$ and $\ell_t$.
On the other hand, the particle trajectory can be kept unchanged while
resetting the boundary local time.  This is a new resetting scheme,
which can model some reactivity dynamics of the target.  Such features
were not available within the conventional description of
diffusion-controlled reactions that focused exclusively on the
position of the particle.  Finally, one can consider even more
sophisticated resetting mechanisms that involve both the position
$\X_t$ of the particle and its boundary local time $\ell_t$.  

The paper is organized as follows.  Section \ref{sec:general} presents
our main results in general bounded domains.  We start in
Sec. \ref{sec:encounter} by recalling the encounter-based approach and
summarizing the main notations and formulas that will be used in the
paper.  In Sec. \ref{sec:position}, we derive the full propagator and
related quantities under resetting of the position, whereas
Sec. \ref{sec:boundary} is devoted to resetting of the boundary local
time.  In Sec. \ref{sec:1D}, we illustrate general results in the case
of diffusion on an interval, for which most quantities can be found
explicitly.  Section \ref{sec:discussion} summarizes our results and
presents further perspectives.

\section{General results}
\label{sec:general}

\subsection{Encounter-based approach}
\label{sec:encounter}

We consider a point-like particle diffusing with a constant
diffusivity $D$ in a bounded Euclidean domain $\Omega\subset \R^d$
with a smooth reflecting boundary $\pa$.  For a particle started at
time $0$ from a point $\x_0 \in \Omega$, the stochastic process $\X_t$
denotes its (random) position at time $t$.  We are interested in
describing surface reactions on a chosen ``target'' $\Gamma$, which is
a subset of the otherwise inert boundary $\pa$.
Following L\'evy, one can introduce the boundary local time $\ell_t$
on $\Gamma$ as \cite{Levy,Ito,Freidlin}
\begin{equation} \label{eq:ell_def}
\ell_t = \lim\limits_{a\to 0} \frac{D}{a} \underbrace{\int\limits_0^t dt'  \, \Theta(a - |\X_{t'} - \Gamma|)}_{\textrm{residence time in}~\Gamma_a} ,
\end{equation}
where $|\x - \Gamma|$ is the Euclidean distance between a point $\x$
and the target set $\Gamma$, and $\Theta(z)$ is the Heaviside step
function: $\Theta(z) = 1$ for $z >0$ and $0$ otherwise.  Here the
integral over $t'$ is the residence time of the particle inside a thin
boundary layer $\Gamma_a = \{ \x\in\Omega ~:~ |\x - \Gamma| < a\}$ of
width $a$ near the target (defined via $\Theta(a - |\X_{t'} -
\Gamma|)$).  As the target $\Gamma$ has a lower dimensionality $d-1$
as compared to the confining domain $\Omega$, the residence time
vanishes in the limit $a\to 0$.  In turn, its rescaling by $a$ yields
a well-defined limit.  Even though $\ell_t$ is called ``boundary local
time'', it has units of length (in turn, $\ell_t/D$ has units of time
per length).  We stress that the mathematical construction of
reflected Brownian motion via the stochastic Skorokhod equation
involves the boundary local time on the whole boundary $\pa$.  Here,
we consider its modified version by restricting the residence time in
Eq. (\ref{eq:ell_def}) to the subset $\Gamma$ which represents a
target.  Accordingly, we introduce the full propagator
$P(\x,\ell,t|\x_0)$ as the joint probability density of the particle
position $\X_t$ inside $\Omega$ and its boundary local time $\ell_t$
{\it on the target}.  We also outline that the {\it boundary} local
time should be distinguished from a {\it point} local time, which
represents the rescaled residence time in a vicinity of a bulk point.
The latter has been intenstively studied (see
\cite{Comtet02,Majumdar02,Majumdar05} and references therein), in
particular, in the context of stochastic resetting \cite{Pal19}.  For
this purpose, one could use the Feynman-Kac formula with
$\delta$-shaped potential in the bulk.  However, the application of
this formula to the boundary local time would require setting singular
$\delta$-potentials on the boundary that renders the whole approach
less appealing (see Sec. E of Supplemental Material of
\cite{Grebenkov20} for further discussions).  For this reason, the
analysis of the boundary local time relies on different mathematical
tools discussed below.

In many physical, chemical and biological applications, the target
$\Gamma$ can be modeled as a thin reactive layer $\Gamma_a$ with some
bulk reactivity $\mu$ (in units 1/s).  When the layer width $a$ is
small, the definition (\ref{eq:ell_def}) indicates that $a\ell_t/D$
characterizes the residence time that a particle has spent in the
reactive layer, up to time $t$.  For a basic first-order reaction
kinetics, the survival probability of the particle $S_q(t|\x_0)$
(i.e., the probability that the particle has not reacted in $\Gamma_a$
up to time $t$) is then
\begin{equation}  \label{eq:Sq_model}
S_q(t|\x_0) = \E \{e^{- \mu a \ell_t/D}\},
\end{equation}
where 
\begin{equation}
q = \frac{\mu a}{D} = \frac{\kappa}{D} \,,
\end{equation}
with $\kappa = \mu a$ being the target reactivity (in units m/s).  By
introducing an independent random variable $\hat{\ell}$ with an
exponential law, $\P\{\hat{\ell} > \ell\} = e^{-q\ell}$, one can
rewrite the above expectation as \cite{Grebenkov20}
\begin{equation}  \label{eq:Sq_rho}
S_q(t|\x_0) = \P\{ \ell_t < \hat{\ell}\} = \int\limits_0^\infty d\ell \, \underbrace{e^{-q\ell}}_{=\P\{\ell < \hat{\ell} \}} \, \rho(\ell,t|\x_0),
\end{equation}
where 
\begin{equation}  \label{eq:rho_def}
\rho(\ell,t|\x_0) = \int\limits_{\Omega} d\x~ P(\x,\ell,t|\x_0) , 
\end{equation}
is the probability density of the boundary local time $\ell_t$ (here
and throughout the text, the real positive variable $\ell$ in
$\rho(\ell,t|\x_0)$ and related expressions denotes any possible
realization of the random variable $\ell_t$).  In this representation,
the reaction on the target $\Gamma$ occurs at the first time when the
boundary local time $\ell_t$ exceeds the random threshold $\hat{\ell}$
that naturally defines the first-reaction time (FRT) $\T$ as
\begin{equation}
\T = \inf\{ t > 0 ~:~ \ell_t > \hat{\ell}\} .
\end{equation}
As the boundary local time $\ell_t$ is a non-decreasing process, one
has $S_q(t|\x_0) = \P\{ \ell_t < \hat{\ell}\} = \P\{ \T > t\}$ so that
the survival probability determines the cumulative distribution
function of the first-reaction time $\T$, its probability density,
\begin{equation} \label{eq:Hq_def}   
H_q(t|\x_0) = -\partial_t S_q(t|\x_0), 
\end{equation}
and all the moments.  Moreover, as the threshold $\hat{\ell}$ is
exponentially distributed, one can express $H_q(t|\x_0)$ as
\begin{equation}  \label{eq:Hq_U}
H_q(t|\x_0) = \int\limits_0^\infty d\ell \, \underbrace{qe^{-q\ell}}_{=\textrm{pdf of~} \hat{\ell}} \, U(\ell,t|\x_0),
\end{equation}
where $U(\ell,t|\x_0)$ is the probability density of the
first-crossing time $\T_\ell$ of a fixed threshold $\ell$ by $\ell_t$:
\begin{equation}  \label{eq:Tell_def}
\T_\ell = \inf\{ t > 0 ~:~ \ell_t > \ell\}.
\end{equation}

In analogy with Eq. (\ref{eq:Sq_rho}), one can relate the full
propagator $P(\x,\ell,t|\x_0)$ to the conventional propagator
$G_q(\x,t|\x_0)$ as \cite{Grebenkov20}
\begin{equation} \label{PtoG}
G_q(\x,t|\x_0) = \int\limits_{0}^{\infty} d\ell \, e^{-q\ell} \, P(\x,\ell,t|\x_0).
\end{equation}
Note that the integral of this relation over $\x\in \Omega$ yields
Eq. (\ref{eq:Sq_rho}).  Here, the full propagator $P(\x,\ell,t|\x_0)$
describes the diffusive dynamics inside the confining domain $\Omega$
with a reflecting inert boundary $\pa$, i.e., {\it without any surface
reaction}, even on the target region $\Gamma$.  In turn, the
conventional propagator $G_q(\x,t|\x_0)$ describes the probability
density of finding the particle at time $t$ in a vicinity of point
$\x$ in the presence of partially reactive target $\Gamma$ with
reactivity parameter $q$, thus accounting for the survival of the
particle.  One sees that the effect of surface reactions on the target
$\Gamma$ is incorporated {\it a posteriori} via the factor
$e^{-q\ell}$.  Moreover, one can replace the exponential law
$e^{-q\ell}$ for the random threshold $\hat{\ell}$ by any law,
$\P\{\hat{\ell} > \ell\} = \Psi(\ell)$, that defines a generalized
propagator
\begin{equation} \label{GPsi}
G_{\Psi}(\x,t|\x_0) = \int\limits_{0}^{\infty} d\ell \, \Psi(\ell) \, P(\x,\ell,t|\x_0),
\end{equation}
and allows one to deal with more sophisticated surface reaction
mechanisms \cite{Grebenkov20}.  

We outline the explicit dependence on $q$ in Eq. (\ref{PtoG}), in
contrast to the conventional descriptions 
\cite{Collins49,Sano79,Sano81,Shoup82,Sapoval94,Filoche99,Benichou00,Sapoval02,Grebenkov03,Grebenkov05,Grebenkov06a,Grebenkov06,Traytak07,Bressloff08,Singer08,Grebenkov10a,Grebenkov10b,Lawley15,Grebenkov15,Chaigneau22}, 
in which the parameter $q$ enters {\it implicitly} through the Robin
boundary condition to the diffusion equation:
\begin{subequations}  \label{eqdiff}
\begin{align} 
 \partial_t G_q(\x,t|\x_0) & = D \Delta G_q(\x,t|\x_0)  \quad (\x\in \Omega),\\   \label{eq:Gq_Robin}
 - \partial_n G_q(\x,t|\x_0) & = q G_q(\x,t|\x_0)    \quad (\x\in \Gamma), \\
 - \partial_n G_q(\x,t|\x_0) & = 0                    \quad (\x\in\pa\backslash\Gamma), 
\end{align}
\end{subequations}
subject to the initial condition $G_q(\x,t=0|\x_0) = \delta(\x-\x_0)$
with the Dirac distribution.  Here $\Delta$ is the Laplace operator
and $\partial_n$ is the normal derivative oriented outwards the domain
$\Omega$.  As stressed in \cite{Grebenkov20}, the Robin boundary
condition (\ref{eq:Gq_Robin}), which is consistent with the
exponential model (\ref{eq:Sq_model}) of the survival probability, is
a {\it choice} of one surface reaction mechanism among many others.

As the domain $\Omega$ is bounded, the Laplace operator has a discrete
spectrum, and the solution of the above equations can be expressed via
a spectral decomposition
\begin{equation}   \label{Gspectral_t}
G_q(\x,t|\x_0) = \sum\limits_{k=0}^\infty u_k^{(q)}(\x) \, [u_k^{(q)}(\x_0)]^* \, e^{-Dt\lambda_k^{(q)}} \,,
\end{equation}
where asterisk denotes complex conjugate, while $\lambda_k^{(q)}$ and
$u_k^{(q)}(\x)$ are the eigenvalues and $L_2(\Omega)$-normalized
eigenfunctions of the (negative) Laplace operator $-\Delta$ in
$\Omega$:
\begin{subequations}  \label{eq:uk}
\begin{align} 
-\Delta u_k^{(q)}(\x) & = \lambda_k^{(q)} \, u_k^{(q)}(\x) \quad (\x\in \Omega),\\   \label{eq:uk_Robin}
- \partial_n u_k^{(q)}(\x) & = q u_k^{(q)}(\x)    \quad (\x\in \Gamma), \\
 - \partial_n u_k^{(q)}(\x) & = 0                    \quad (\x\in\pa\backslash\Gamma).
\end{align}
\end{subequations}
The spectral expansion (\ref{Gspectral_t}) highlights the symmetry of
the propagator with respect to the exchange of the starting and
arrival points $\x_0$ and $\x$: $G_q(\x,t|\x_0) = G_q(\x_0,t|\x)$.
The spectral expansion of the full propagator, derived in
\cite{Grebenkov20}, implies the same symmetry 
\begin{equation}  \label{eq:Psymmetry}
P(\x,\ell,t|\x_0) = P(\x_0,\ell,t|\x).
\end{equation}
Such a symmetry can be broken in the presence of a drift or a
potential (see \cite{Grebenkov22} for details).

In the following, we describe how resetting may affect the above
quantities: the full propagator $P(\x,\ell,t|\x_0)$, the conventional
propagator $G_q(\x,t|\x_0)$, and the probability densities
$\rho(\ell,t|\x_0)$, $H_q(t|\x_0)$ and $U(\ell,t|\x_0)$ of the
boundary local time $\ell_t$, of the first-reaction time $\T$, and of
the first-crossing time $\T_\ell$, respectively.

\subsection{Position resetting}
\label{sec:position}

We start by looking at the conventional scenario of stochastic
resetting of the particle position.  At each resetting, the particle
is immediately relocated to its starting position $\x_0$.  We assume
that durations $\delta_1$, $\delta_2$, etc.  between consecutive
resettings are independent identically distributed random variables
drawn from a prescribed probability density function (PDF) $\phi(t)$.
In other words, resettings occur at random times $t_1 = \delta_1, ~t_2
= \delta_1 + \delta_2, ~\ldots,~ t_k = \delta_1 + \ldots + \delta_k$,
etc.  We first present general results and then discuss the Poissonian
resetting with a rate $\omega$ (i.e., $\phi(t) = \omega e^{-\omega
t}$) as one of the most common models of stochastic resetting.  Note
that the Poissonian case was recently studied in
\cite{Bressloff22c} (see also Sec. \ref{sec:discussion}).

\subsubsection*{General results}

Following and extending the renewal approach from \cite{Chechkin18}
(see also \cite{Evans19,Evans20,Ross}), we compute the full propagator
with resetting, denoted as $P_\phi(\x,\ell,t| \x_0)$, by counting the
number of resettings up to time $t$ and adding their contributions:
\begin{widetext}
\begin{align}  \nonumber
P_\phi(\x,\ell,t| \x_0) & = \Phi(t) P(\x,\ell,t|\x_0) + \int\limits_{0}^{t} dt_1 \phi(t_1) \int\limits_{\Omega} \int\limits_{0}^{\ell} 
d\x_1 d\ell_1 P(\x_1,\ell_1,t_1|\x_0) \, \Phi(t-t_1) P(\x,\ell-\ell_1,t-t_1|\x_0) \\   \nonumber
& +  \int\limits_{0}^{t} dt_1 \phi(t_1) \int\limits_{\Omega}  \int\limits_{0}^{\ell} d\x_1 d\ell_1 P(\x_1,\ell_1,t_1|\x_0)
 \int\limits_{0}^{t-t_1} dt_2 \phi(t_2) \int\limits_{\Omega} \int\limits_{0}^{\ell-\ell_1} d\x_2  d\ell_2  P(\x_2,\ell_2,t_2|\x_0) \\ 
\label{eqmaitresse_Pphi}
& \times \Phi(t-t_1-t_2) P(\x,\ell-\ell_1-\ell_2,t-t_1-t_2|\x_0) +  \ldots  ,
\end{align}
\end{widetext}
where $ \Phi(t) = \int\nolimits_{t}^{\infty} dt' \phi(t')$ is the
probability of no resetting up to time $t$.
The first term is the contribution without resetting.  The second term
describes one resetting at time $t_1$ (with probability
$\phi(t_1)dt_1$), which can range from $0$ to $t$.  The factor
$P(\x_1,\ell_1,t_1|\x_0) d\x_1 d\ell_1$ describes the probability for
the particle to be at an intermediate position $\x_1$ with an
intermediate boundary local time $\ell_1$.  After resetting, the
position of the particle is reset to $\x_0$, while the boundary local
time remains unchanged.  The Markov property implies that the
remaining part of the diffusive process, from $t_1$ to $t$, is
described by the probability density $P(\x,\ell-\ell_1,t-t_1|\x_0)$,
while $\Phi(t-t_1)$ ensures that there is no resetting during that
period.  Similarly, the third, fourth, etc. terms describe the
contributions of 2, 3, etc. resettings.  Note that one can easily
implement the case when the resetting position is different from the
starting point $\x_0$.

Using the probability density $\rho(\ell,t|\x_0)$ of the boundary
local time $\ell_t$ defined by Eq. (\ref{eq:rho_def}), one can
simplify the integrals over intermediate positions $\x_1$, $\x_2$,
etc.  In turn, the convolutions over $\ell_k$ and $t_k$ can be turned
into products by performing the double Laplace transform with respect
to variables $\ell$ and $t$ that is defined for a given function
$f(\ell,t)$ as
\begin{equation}
\L_{q,p} \bigl\{ f(\ell,t) \bigr\} = 
\int\limits_0^\infty d\ell \, e^{-q\ell} \int\limits_0^\infty dt \, e^{-pt} \, f(\ell,t) .
\end{equation}
Applying this transform to Eq. (\ref{eqmaitresse_Pphi}) and summing
the resulting geometric series, one has
\begin{equation} \label{eqfinale_Pphi0}
\L_{q,p} \bigl\{P_\phi(\x,\ell,t|\x_0) \bigr\} = \frac{\L_{q,p} \bigl\{\Phi(t) P(\x,\ell,t|\x_0) \bigr\}}
{1 - \L_{q,p} \bigl\{\phi(t) \rho(\ell,t|\x_0) \bigr\}} \,.
\end{equation}
As $\Phi(t)$ and $\phi(t)$ do not depend on $\ell$, one can first
perform the Laplace transform with respect to $\ell$ to get
\begin{equation} \label{eqfinale_Pphi}
\L_{q,p} \bigl\{P_\phi(\x,\ell,t|\x_0) \bigr\} = \frac{\L_{p} \bigl\{\Phi(t) G_q(\x,t|\x_0) \bigr\}}
{1 - \L_{p} \bigl\{\phi(t) S_q(t|\x_0) \bigr\}} \,,
\end{equation}
where we used Eqs. (\ref{eq:Sq_rho}, \ref{PtoG}), and $\L_p$ denotes
the Laplace transform with respect to $t$.  An inversion of the double
Laplace transform in this equation formally yields the full propagator
$P_\phi(\x,\ell,t|\x_0)$ under resetting:
\begin{equation}   \label{eq:Pphi_t}
P_\phi(\x,\ell,t|\x_0) = \L_{q,p}^{-1} \biggl\{\frac{\L_{p} \bigl\{\Phi(t) G_q(\x,t|\x_0) \bigr\}}
{1 - \L_{p} \bigl\{\phi(t) S_q(t|\x_0) \bigr\}} \biggr\} \,,
\end{equation}
where $\L_{q,p}^{-1}$ denotes the double inverse Laplace transform
with respect to $q$ and $p$.
This expression outlines that position resetting breaks the symmetry
of the full propagator with respect to the exchange between $\x$ and
$\x_0$, in sharp contrast to Eq. (\ref{eq:Psymmetry}) for the full
propagator $P(\x,\ell,t|\x_0)$ without resetting.  This is not
surprising because resetting to the starting point $\x_0$
distinguishes it from other points.  Expectedly, the form of
Eq. (\ref{eqfinale_Pphi}) resembles Eq. (1) from \cite{Chechkin18} for
the probability density of the first-passage time.  However,
Eq. (\ref{eqfinale_Pphi}) gives access to more refined information
about the diffusing particle -- to its full propagator under
resetting.

In analogy with Eq. (\ref{PtoG}), the Laplace transform of
$P_\phi(\x,\ell,t|\x_0)$ with respect to $\ell$ determines the
conventional propagator under resetting that we denote as
$G_{q,\phi}(\x,t|\x_0)$.  In other words, Eq. (\ref{eqfinale_Pphi})
can be written as
\begin{equation} \label{eqfinale_Pphi2}
\tilde{G}_{q,\phi}(\x,p|\x_0) = \frac{\L_{p} \bigl\{\Phi(t) G_q(\x,t|\x_0) \bigr\}}{1 - \L_{p} \bigl\{\phi(t) S_q(t|\x_0) \bigr\}} \,,
\end{equation}
where tilde denotes the Laplace transform $\L_p$ of
$G_{q,\phi}(\x,t|\x_0)$ with respect to time $t$ (we keep using this
tilde notation for other quantities in the following).  The inverse
Laplace transform formally yields the propagator under resetting in
time domain:
\begin{equation}  \label{eqfinale_Pphi2_bis}
G_{q,\phi}(\x,t|\x_0) = \L_{p}^{-1} \biggl\{\frac{\L_{p} \bigl\{\Phi(t) G_q(\x,t|\x_0) \bigr\}}{1 - \L_{p} \bigl\{\phi(t) S_q(t|\x_0) 
\bigr\}} \biggr\}\,.
\end{equation}

For a reactive target ($q > 0$), the propagator
$G_{q,\phi}(\x,t|\x_0)$ is expected to vanish in the long-time limit,
as the particle diffusing in a bounded domain cannot in general avoid
hitting the target (see the discussion below in the case of a
Poissonian resetting).  In turn, if the target is inert ($q = 0$), the
particle survives forever, $S_0(t|\x_0) = 1$, whereas the conventional
propagator approaches a steady-state uniform distribution:
$G_0(\x,t|\x_0) \to 1/|\Omega|$ as $t\to\infty$, where $|\Omega|$ is
the volume of the confining domain $\Omega$.  If the resetting time
density $\phi(t)$ has a finite mean $\E\{ \delta\}$, the denominator
of Eq. (\ref{eqfinale_Pphi2_bis}) behaves in the small-$p$ limit as
$1 - \L_{p} \bigl\{\phi(t) S_0(t|\x_0) \bigr\} = 1 - \tilde{\phi}(p) \approx p \E\{\delta\} + O(p^2)$,
that implies the long-time behavior
\begin{equation} \label{eq:Gt_asympt_q0}
G_{0,\phi}(\x,t|\x_0) \xrightarrow{t\to\infty} 
G_{0,\phi}^{\rm st}(\x|\x_0) = \int\limits_0^\infty \frac{dt \, \Phi(t)}{\E\{ \delta\}}  G_0(\x,t|\x_0) \,.
\end{equation}
When there was no resetting, the diffusing particle explored the
bounded confining domain and therefore equilibrated the likelihood of
its location at any point in $\Omega$ (the uniform steady-state
distribution).  Moreover, the information on the starting point was
lost.  In contrast, resetting breaks this uniformity and preserves
information on the resetting point in the steady-state distribution
$G_{0,\phi}^{\rm st}(\x|\x_0)$.

The integral of the propagator in Eq. (\ref{eqfinale_Pphi2_bis}) over
$\x\in\Omega$ determines the survival probability under resetting:
\begin{equation} \label{eq:Sphi0}
S_{q,\phi}(t|\x_0) = \L_{p}^{-1} \biggl\{ \frac{\L_{p} \bigl\{\Phi(t) S_q(t|\x_0) \bigr\}}
{1 - \L_{p} \bigl\{\phi(t) S_q(t|\x_0) \bigr\}} \biggr\} \,,
\end{equation}
or, equivalently,
\begin{equation}  \label{eq:Sphi}
\tilde{S}_{q,\phi}(p|\x_0) = \frac{\L_{p} \bigl\{\Phi(t) S_q(t|\x_0) \bigr\}}
{1 - \L_{p} \bigl\{\phi(t) S_q(t|\x_0) \bigr\}} \,.
\end{equation}
As previously, the survival probability determines the probability
density $H_{q,\phi}(t|\x_0) = -\partial_t S_{q,\phi}(t|\x_0)$ of the
first-reaction time $\T_\phi$ under resetting, as well as its moments.
In particular, the mean FRT under resetting is
\begin{align}  \nonumber
\E\{ \T_{\phi} \} & = \int\limits_0^\infty dt \, t \, H_{q,\phi}(t|\x_0) = \int\limits_0^\infty dt \, S_{q,\phi}(t|\x_0) \\  \label{eq:Tphi_general}
& = \tilde{S}_{q,\phi}(0|\x_0) = \frac{\int\nolimits_0^\infty dt \, \Phi(t) \, S_q(t|\x_0)}{\int\nolimits_0^\infty dt \, \Phi(t) \, H_q(t|\x_0)} \,,
\end{align}
where we used that $\phi(t) = -\partial_t \Phi(t)$,
Eq. (\ref{eq:Hq_def}), and integrated by parts in the denominator.
This expression was earlier derived by Pal {\it et al.} \cite{Pal16},
as well as by Chechkin and Sokolov \cite{Chechkin18}, who used it to
investigate the search optimality under resetting (see also
\cite{Evans20}).  Note that Eq. (\ref{eq:Tphi_general}) is equivalent
to the general form (\ref{eq:T_Pal}) obtained by Pal and Reuveni
\cite{Pal17}.  We also note that the relation (\ref{eq:Hq_U}), which
is valid under resetting, can be inverted to access the probability
density of the first-crossing time under resetting:
\begin{equation} \label{eq:U_phi}
U_\phi(\ell,t|\x_0) = \L_{q,p}^{-1} \biggl\{ \frac{\tilde{H}_{q,\phi}(p|\x_0)}{q} \biggr\} \,.
\end{equation}
Finally, the integral of Eq. (\ref{eq:Pphi_t}) over
$\x\in\Omega$ yields the probability density $\rho_\phi(\ell,t|\x_0)$
of the boundary local time under resetting:
\begin{equation}    \label{eq:rho_phi_t}
\rho_\phi(\ell,t|\x_0) = \L_{q,p}^{-1} \biggl\{\frac{\L_{p} \bigl\{\Phi(t) S_q(t|\x_0) \bigr\}}
{1 - \L_{p} \bigl\{\phi(t) S_q(t|\x_0) \bigr\}} \biggr\} \,.
\end{equation}

\subsubsection*{Poissonian resetting}

For the Poissonian resetting with $\Phi(t) = e^{-\omega t}$, we use
again Eq. (\ref{PtoG}) to simplify Eq. (\ref{eqfinale_Pphi2}) as
\begin{equation}  \label{eqfinale_Gw}
\tilde{G}_{q,\omega}(\x,p|\x_0) = \frac{\tilde{G}_q(\x,p+\omega|\x_0)}{1 - \omega \tilde{S}_q(p+\omega|\x_0)} \,.
\end{equation}
One can invert this Laplace transform via the residue theorem by
finding the poles $\{p_k\} \subset \C$ of
$\tilde{G}_{q,\omega}(\x,p|\x_0)$.  For $q > 0$, these poles are
determined by the equation:
\begin{equation}  \label{eq:poles}
\tilde{S}_q(p_k+\omega|\x_0) = \frac{1}{\omega} \,.
\end{equation}
In the limit $\omega \to 0$, the resetting is progressively switched
off, and the $k$-th pole $p_k$ approaches $-D\lambda_k^{(q)}$.  The
pole $p_0$ with the largest real part determines the exponential decay
of the propagator $G_{q,\omega}(\x,t|\x_0)$ in the long-time limit.

In turn, for the inert target ($q = 0$), Eq. (\ref{eqfinale_Gw})
simplifies to
\begin{equation}
\tilde{G}_{0,\omega}(\x,p|\x_0) = \biggl(1 + \frac{\omega}{p}\biggr) \tilde{G}_0(\x,p+\omega|\x_0) \,,
\end{equation}
which in time domain reads
\begin{align}  \nonumber
G_{0,\omega}(\x,t|\x_0) & = e^{-\omega t} G_0(\x,t|\x_0) \\
& + \omega \int\limits_0^t dt' e^{-\omega t'} G_0(\x,t'|\x_0) \,.
\end{align}
This renewal-type equation has been used in many earlier works (see
\cite{Evans20} and references therein).  In the limit $t\to\infty$,
one gets
\begin{equation} \label{eq:Gt_asympt_q0_exp}
G_{0,\phi}(\x,t|\x_0) \xrightarrow{t\to\infty} \omega \tilde{G}_0(\x,\omega|\x_0) \,,
\end{equation}
in agreement with Eq. (\ref{eq:Gt_asympt_q0}).

Other general expressions are also simplified for the Poissonian
resetting.  For instance, Eq. (\ref{eq:Sphi}) reads
\begin{equation}  \label{eqfinale_Sw}
\L_{q,p}\{\rho_{\omega}(\ell,t|\x_0)\} = \tilde{S}_{q,\omega}(p|\x_0) 
= \frac{\tilde{S}_q(p+\omega|\x_0)}{1 - \omega \tilde{S}_q(p+\omega|\x_0)} \,,
\end{equation}
from which the probability density follows as
\begin{equation}  \label{eqfinale_Hw}
\tilde{H}_{q,\omega}(p|\x_0) = \frac{(p+\omega) \tilde{H}_q(p+\omega|\x_0)}{p + \omega \tilde{H}_q(p+\omega|\x_0)} \,.
\end{equation}  
This relation was reported by Reuveni and used to show that the
relative standard deviation of the FPT is equal to $1$ under optimal
resetting \cite{Reuveni16} (see also \cite{Bressloff22c}; 
a similar relation for the generating function was given in
\cite{Kusmierz14}; see a more general discussion in Sec. 3.1 of the
review \cite{Evans20}).  In particular, setting $p = 0$ in
Eq. (\ref{eqfinale_Sw}) yields the known expression for the mean FRT:
\begin{equation}  \label{eq:Tphi_omega}
\E\{ \T_{\omega} \} = \frac{\tilde{S}_q(\omega|\x_0)}{1 - \omega \tilde{S}_q(\omega|\x_0)} 
= \frac{1 - \tilde{H}_q(\omega|\x_0)}{\omega \tilde{H}_q(\omega|\x_0)} \,.
\end{equation}
Expressing $\tilde{S}_q(p+\omega|\x_0)$ in terms of
$\tilde{S}_{q,\omega}(p|\x_0)$ from Eq. (\ref{eqfinale_Sw}) and then
substituting it into Eq. (\ref{eqfinale_Gw}) gives 
\begin{equation}
\tilde{G}_{q,\omega}(\x,p|\x_0) = \bigl(1 + \omega \tilde{S}_{q,\omega}(p|\x_0)\bigr) \tilde{G}_{q}(\x,p+\omega|\x_0) ,
\end{equation}
from which the double inverse Laplace transform with respect to $p$
and $q$ yields
\begin{align}   \label{eq:Pomega_position}
& P_{\omega}(\x,\ell,t|\x_0) = e^{-\omega t} P(\x,\ell,t|\x_0) \\  \nonumber
& + \int\limits_0^t dt' \omega e^{-\omega (t-t')} \hspace*{-1mm} \int\limits_0^\ell d\ell' P(\x,\ell-\ell',t-t'|\x_0) \, \rho_\omega(\ell',t'|\x_0).
\end{align}  
This is a typical renewal-type representation, in which the first term
represents the contribution without resetting, while the second term
accounts for resettings; in this term, the time interval from $0$ to
$t$ is split by time $t'$ of the {\it last} resetting before $t$,
which occurs with the probability density $\omega e^{-\omega (t-t')}$.
During the period from $0$ to $t'$, resettings erase information on
the position so that $\rho_\omega(\ell',t'|\x_0)$ determines the
boundary local time $\ell'$ acquired up to $t'$.  In turn, as the
position is reset to $\x_0$ at $t'$, the diffusive dynamics from $t'$
to $t$ is described by $P(\x,\ell-\ell',t-t'|\x_0)$.
The relation (\ref{eq:Pomega_position}), which was also derived in
\cite{Bressloff22c}, expresses the full propagator under Poissonian
resetting in terms of $P(\x,\ell,t|\x_0)$ (without resetting) and the
probability density $\rho_\omega(\ell,t|\x_0)$ with resetting.
However, its explicit form is deceptive because
$\rho_\omega(\ell,t|\x_0)$ still has to be determined via the double
inverse Laplace transform of Eq. (\ref{eqfinale_Sw}).

\subsection{Boundary local time resetting}
\label{sec:boundary}

Now we turn to another resetting scenario, which was not studied
earlier and consists in resetting the boundary local time $\ell_t$,
while keeping the position $\X_t$ unchanged.  Such a resetting does
not affect the dynamics of the particle; in particular, if the target
is inert ($q = 0$), the related propagator $G_0(\x,t|\x_0)$ remains
unchanged by construction.  However, resetting of the boundary local
time may affect the reaction mechanism on a reactive target.  We aim
therefore to analyze how such a resetting modifies the full propagator
$P(\x,\ell,t|\x_0)$, the propagator $G_q(\x,t|\x_0)$, and the
probability density $\rho(\ell,t|\x_0)$ of the boundary local time.
We also discuss the (unsolved) challenges in computing the probability
density $U(\ell,t|\x_0)$ of the first-crossing time.

\subsubsection*{Full propagator}

Similarly to Eq. (\ref{eqmaitresse_Pphi}), one can write a
renewal-type relation
\begin{widetext}
\begin{equation} \label{eqmaitresse_Pw_l}
P_\phi(\x,\ell,t|\x_0) =  \Phi(t) P(\x,\ell,t|\x_0) + \int\limits_{0}^{t} dt_1 \phi(t_1) 
\int\limits_{\Omega}  \int\limits_{0}^{\infty} d\x_1 d\ell_1  P(x_1,\ell_1,t_1|\x_0) \, \Phi(t-t_1) P(\x,\ell,t-t_1|\x_1)   +  \ldots   
\end{equation}
\end{widetext}
In contrast to the previous computation in Sec. \ref{sec:position},
convolutions over boundary local times are replaced by integrals over
their intermediate values.  Using Eq. (\ref{PtoG}) with $q = 0$ to
evaluate these integrals and performing the Laplace transform with
respect to time $t$, we get
\begin{align}  \label{eq:Phi_phi_auxil1}
& \tilde{P}_\phi(\x,\ell,p| \x_0) = \L_p\{ \Phi(t) P(\x,\ell,t|\x_0)\} \\  \nonumber  
& + \int\limits_{\Omega}  d\x_1 \L_p\{ \phi(t) G_0(\x_1,t|\x_0)\} \L_p\{ \Phi(t) P(\x,\ell,t|\x_1)\} + \ldots 
\end{align}
To proceed, one needs to evaluate the integrals over intermediate
positions $\x_k \in \Omega$.  For this purpose, we use the spectral
decomposition (\ref{Gspectral_t}) of the propagator $G_0(\x,t|\x_0)$.
For instance, one has
\begin{equation}
\L_p\{ \phi(t) G_0(\x,t|\x_0)\} = \sum\limits_{k=0}^\infty u_k^{(0)}(\x)  [u_k^{(0)}(\x_0)]^*  \tilde{\phi}(p+D\lambda_k^{(0)}) .
\end{equation}
The orthonormality of eigenfunctions $u_k^{(0)}(\x)$ allows one to
compute the integral of two such functions:
\begin{align*}   \nonumber
& \int\limits_{\Omega}  d\x_1 \, \L_p\{ \phi(t) G_0(\x_1,t|\x_0)\} \, \L_p\{ \phi(t) G_0(\x,t|\x_1)\}  \\  \nonumber
& = \sum\limits_{k_1,k_2=0}^\infty \int\limits_{\Omega}  d\x_1 \, \biggl(u_{k_1}^{(0)}(\x_1) \, [u_{k_1}^{(0)}(\x_0)]^* \, 
\tilde{\phi}(p+D\lambda_{k_1}^{(0)})\biggr) \\  \nonumber
& \times  \biggl( u_{k_2}^{(0)}(\x) \, [u_{k_2}^{(0)}(\x_1)]^* \, \tilde{\phi}(p+D\lambda_{k_2}^{(0)})\biggr) \\  
& = \sum\limits_{k=0}^\infty u_{k}^{(0)}(\x) \, [u_{k}^{(0)}(\x_0)]^* \, \biggl[\tilde{\phi}(p+D\lambda_{k}^{(0)})\biggr]^2 .
\end{align*}
More generally, the integral of over $n-1$ intermediate points $\x_1,
\x_2, \ldots,\x_{n-1}$ yields the $n$-th power of
$\tilde{\phi}(p+D\lambda_{k}^{(0)})$.  Using this property, we can sum
the infinite number of terms in Eq. (\ref{eq:Phi_phi_auxil1}) to get
\begin{align}  \nonumber
\tilde{P}_\phi(\x,\ell,p| \x_0) & = \int\limits_{\Omega}  d\x' \L_p\{ \Phi(t) P(\x,\ell,t|\x')\} \\  \label{eq:Phi_phi_general}
& \times \sum\limits_{k=0}^\infty  \frac{u_k^{(0)}(\x') [u_k^{(0)}(\x_0)]^*}{1 - \tilde{\phi}(p+D\lambda_{k}^{(0)})} \,.
\end{align}
An inverse Laplace transform with respect to $p$ yields a formal
solution for the full propagator $P_\phi(\x,\ell,t| \x_0)$ under
resetting of the boundary local time:
\begin{align} \nonumber
P_\phi(\x,\ell,t| \x_0) & = \int\limits_{\Omega}  d\x' \int\limits_0^t dt' \, \xi(t') \, G_0(\x',t'|\x_0)    \\  \label{eq:Phi_phi_general_t}
& \times   \Phi(t-t') \, P(\x,\ell,t-t'|\x') ,  
\end{align}
where
\begin{equation}
\xi(t) = \L_p^{-1} \biggl\{ \frac{1}{1 - \tilde{\phi}(p)} \biggr\}(t),
\end{equation}
and we used again the spectral decomposition (\ref{Gspectral_t}) of
the propagator $G_0(\x,t|\x_0)$.

It is instructive to look at the long-time behavior of the full
propagator.  As the position of the particle is not affected by
resetting, it should reach the uniform distribution, as in the
no-resetting case.  In addition, random resettings of the boundary
local time render this quantity stationary at long times as well.  As
a consequence, one can expect that the full propagator reaches a
well-defined steady-state limit.  This is indeed the case.  To show
it, let us examine Eq. (\ref{eq:Phi_phi_general}) in the limit $p\to
0$, which corresponds to the long-time behavior.  It is known that the
principal eigenvalue $\lambda_0^{(0)}$ of the Laplace operator with
Neumann boundary condition ($q = 0$) is zero.  In addition, the
corresponding eigenfunction is constant: $u_0^{(0)} =
|\Omega|^{-1/2}$.  As a consequence, the sum in
Eq. (\ref{eq:Phi_phi_general}) behaves as $1/(p|\Omega| \E\{\delta\})
+ O(1)$ as $p\to 0$, where we used $\tilde{\phi}(p) \approx 1 -
p\E\{\delta\} + O(p^2)$, under the assumption that the mean
$\E\{\delta\}$ exists.  One sees that the right-hand side of
Eq. (\ref{eq:Phi_phi_general}) has a pole at $p = 0$ that yields the
constant term in the long-time limit:
\begin{equation}  \label{eq:Pphi_asympt}
P_\phi(\x,\ell,t| \x_0) \xrightarrow{t\to\infty} P_{\phi}^{\rm st}(\x,\ell) =  
\int\limits_0^\infty \frac{dt \, \Phi(t)}{\E\{\delta\}}  P(\x,\ell,t|\circ),
\end{equation}
where 
\begin{equation}
P(\x,\ell,t|\circ) = \frac{1}{|\Omega|} \int\limits_{\Omega} d\x' \, P(\x,\ell,t|\x')
\end{equation}
can be interpreted as the full propagator averaged over the starting
point uniformly distributed in $\Omega$ (here $\circ$ highlights that
the starting point is uniformly distributed; we keep using this
notation for other quantities).  Expectedly, the steady-state
distribution $P_{\phi}^{\rm st}(\x,\ell)$ does not depend on the
starting point $\x_0$.

Let us have a closer look at the steady-state limit $P_{\phi}^{\rm
st}(\x,\ell)$.  On one hand, its integral over $\ell$ yields the
expected uniform distribution of the position:
\begin{equation}
\int\limits_0^\infty d\ell \, P_{\phi}^{\rm st}(\x,\ell) = \frac{1}{\E\{\delta\}} \int\limits_0^\infty dt \, \Phi(t) G_0(\x,t|\circ) 
= \frac{1}{|\Omega|} \,,
\end{equation}
where we used that $G_0(\x,t|\circ) = 1/|\Omega|$ for any time $t$ (if
the initial distribution was uniform, it remains uniform for any $t$
since the target is inert).  On the other hand, the joint steady-state
probability density $P_{\phi}^{\rm st}(\x,\ell)$ is not factored,
revealing correlations between $\X_t$ and $\ell_t$.  The steady-state
probability density of the boundary local time reads
\begin{equation}  \label{eq:rho_steady}
\rho^{\rm st}_{\phi}(\ell) = \int\limits_{\Omega} d\x \, P_{\phi}^{\rm st}(\x,\ell)
= \frac{1}{\E\{\delta\}} \int\limits_0^\infty dt \, \Phi(t) \rho(\ell,t|\circ),
\end{equation}
where
\begin{equation}
\rho(\ell, t| \circ) = \frac{1}{|\Omega|} \int\limits_{\Omega} d\x_0 \, \rho(\ell, t | \x_0).
\end{equation}

The expression (\ref{eq:Phi_phi_general}) can be further simplified if
the starting point $\x_0$ is not fixed but uniformly distributed in
$\Omega$.  The orthogonality of eigenfunctions $u_k^{(0)}$ with $k >
0$ to $u_0^{(0)} = |\Omega|^{-1/2}$ implies that the integral over
$\x_0$ cancels all terms in the sum except $k = 0$.  After
simplifications, we get
\begin{align}  \nonumber
\tilde{P}_\phi(\x,\ell,p| \circ) &= \frac{1}{|\Omega|} \int\limits_{\Omega}  d\x_0 \, \tilde{P}_\phi(\x,\ell,p| \x_0) \\  \label{eq:Phi_phi_general_circ}
& = \frac{\L_p\{ \Phi(t) P(\x,\ell,t|\circ)\}}{1 - \tilde{\phi}(p)}   \,.
\end{align}

For the Poissonian resetting, one has $\tilde{\phi}(p) =
\omega/(\omega + p)$ that allows one to simplify
Eq. (\ref{eq:Phi_phi_general}) as
\begin{align}  \nonumber
& \tilde{P}_\omega(\x,\ell,p| \x_0) = \tilde{P}(\x,\ell,p+\omega|\x_0) \\    \label{eqfinale_Pw_l}
& \qquad +  \omega  \int\limits_{\Omega} d\x' \, \tilde{G}_0(\x',p|\x_0) \, \tilde{P}(\x,\ell,p+\omega|\x'),
\end{align}
which reads in time domain as
\begin{align} \label{eqfinale_Pw_l2}
& P_\omega(\x,\ell,t| \x_0) = e^{-\omega t} P(\x,\ell,t|\x_0) \\     \nonumber
& +   \int\limits_0^t dt' \omega e^{-\omega (t-t')} \int\limits_{\Omega} d\x' \, G_0(\x',t'|\x_0) \,  P(\x,\ell,t-t'|\x').
\end{align}
This relation has a simple probabilistic interpretation in terms of
the last resetting time, in analogy to the discussion after
Eq. (\ref{eq:Pomega_position}).  It could also be directly deduced
from Eq. (\ref{eq:Phi_phi_general_t}).

If the starting point $\x_0$ is distributed uniformly, the volume
average of Eq. (\ref{eqfinale_Pw_l}) yields
\begin{equation}  \label{eq:tildePomega_circ}
\tilde{P}_\omega(\x,\ell,p| \circ) = \tilde{P}(\x,\ell,p+\omega|\circ) \biggl(1 + \frac{\omega}{p}\biggr),
\end{equation}
where we used that
\begin{equation}
\int\limits_{\Omega} d\x_{0} \tilde{G}_{0}(\x' , p |\x_{0}) = \int\limits_{\Omega} d\x_{0} \tilde{G}_{0}(\x_{0}, p |\x') = \frac{1}{p}
\end{equation}
due to the normalization of the propagator $G_0(\x_0,t|\x')$ and its
symmetry with respect to the exchange of the starting and arrival
points.  In time domain, Eq. (\ref{eq:tildePomega_circ}) reads
\begin{equation} \label{eq:Pomega_circ}
P_{\omega}(\x,\ell, t | \circ) = e^{-\omega t} P(\x,\ell, t |\circ) +  \int\limits_{0}^{t} dt' \omega e^{-\omega t'} 
P(\x,\ell, t' |\circ).
\end{equation}
In the long-time limit, one retrieves
\begin{equation}  \label{eq:Pomega_circ_steady}
P_{\omega}(\x,\ell, t | \circ) \xrightarrow{t\to\infty} P_{\omega}^{\rm st}(\x,\ell) = \omega \tilde{P}(\x,\ell,\omega|\circ),
\end{equation}
in agreement with the general relation (\ref{eq:Pphi_asympt}).

The full propagator in Eq. (\ref{eq:Phi_phi_general}) determines the
probability density $\rho_{\phi}(\ell,t|\x_0)$ of the boundary local
time under resetting.  In the case of Poissonian resetting, the
integral of Eq. (\ref{eqfinale_Pw_l}) over $\x\in\Omega$ gives
\begin{align}  \nonumber
& \tilde{\rho}_{\omega}(\ell, p | \x_{0}) = \tilde{\rho}(\ell, p+\omega | \x_{0}) \\   \label{eqmaitresse_rhow_l}
& \qquad + \omega \int\limits_{\Omega} d\x' \, \tilde{G}_{0}(\x' , p | \x_{0}) \, \tilde{\rho}(\ell, p+\omega | \x').
\end{align}
If the starting point $\x_0$ is distributed uniformly in $\Omega$, the
average over $\x_0$ yields 
\begin{equation}  \label{eqfinale_rhow_l}
\tilde{\rho}_{\omega}(\ell, p | \circ) = \tilde{\rho}(\ell, p+\omega | \circ) \left(1+\frac{\omega}{p}\right),
\end{equation}
which becomes in time domain
\begin{equation} \label{eqfinale_rhow_l_temporel}
\rho_{\omega}(\ell, t | \circ) = e^{-\omega t} \rho(\ell, t |\circ) +  \int\limits_{0}^{t} dt'\, \omega e^{-\omega t'}\, 
\rho(\ell, t' |\circ),
\end{equation}
in analogy with Eq. (\ref{eq:Pomega_circ}).

In the long-time limit, one retrieves
\begin{equation}  \label{eq:rho_omega_steady}
\rho_{\omega}(\ell, t | \circ) \xrightarrow{t\to\infty} \rho_{\omega}^{\rm st}(\ell) = \omega \tilde{\rho}(\ell,\omega|\circ),
\end{equation}
in agreement with the general relation (\ref{eq:rho_steady}).

To complete this discussion, let us check whether the symmetry of the
full propagator under the exchange of the starting and arrival points
$\x_0$ and $\x$ is preserved or not.  This was the case for the full
propagator $P(\x,\ell,t|\x_0)$ without resetting, see
Eq. (\ref{eq:Psymmetry}).  As resettings of the boundary local time do
not affect the position $\X_t$, one might think that the symmetry is
preserved here.  However, an inspection of
Eq. (\ref{eq:Phi_phi_general}) reveals that this symmetry is in
general broken for the full propagator $P_\phi(\x,\ell,t| \x_0)$ under
resetting.  What does break the symmetry?  Even though the random
trajectory $\X_t$ of the diffusing particle remains unaffected,
resettings modify the boundary local time $\ell_t$ and thus affect
{\it correlations} between $\X_t$ and $\ell_t$, which are captured by
the full propagator $P_\phi(\x,\ell,t|\x_0)$.  To better illustrate
this point, let us consider the situation with a single resetting at
time $t_1$.  The path from $\x_0$ to $\x$ is then split into two
parts: the path from $\x_0$ to an intermediate position $\X_{t_1} =
\x_1$, which is sampled without any constraint on the intermediate
boundary local time $\ell_1$ (ranging from $0$ to $\infty$), and the
path from $\x_1$ to $\x$ with the constraint on $\ell_t$ to be $\ell$.
The reversed path from $\x$ to $\x_0$ (after the exchange of the
starting and arrival points) is also split into similar two parts:
from $\x$ to $\x_1$ without constraint on the boundary local time, and
from $\x_1$ to $\x_0$ with such a constraint.  The presence of a
constraint changes the statistics and thus breaks the symmetry.  This
is particularly clear from Eq. (\ref{eqfinale_Pw_l}) for the
Poissonian resetting: the part without constraint is sampled with
$\tilde{G}_0(\x_1,p|\x_0)$, whereas the part with constraint is
sampled with $\tilde{P}(\x,\ell,p+\omega|\x_1)$.

\subsubsection*{Conventional propagator}

In contrast to Eq. (\ref{PtoG}) in the no-resetting case, the full
propagator $P_\phi(\x,\ell,t|\x_0)$ under resetting, given by
Eq. (\ref{eq:Phi_phi_general}), does not allow one to access the
propagator $G_{q,\phi}(\x,t|\x_0)$.  In fact, one cannot use anymore
the fundamental relation (\ref{PtoG}) because the monotonous growth of
the boundary local time is broken by resettings.  Even if the survival
probability of the particle is still determined as $\E\{e^{-\mu
T_t}\}$ by the total residence time $T_t$ of the particle in a
vicinity of the target, the latter is not given by $a
\ell_t/D$ but should include the boundary local times acquired upon
all prior resettings.  If there were $k$ resettings up to time $t$ at
times $t_1, \ldots, t_k$, then
\begin{equation}
T_t = \frac{a}{D} L_t, \qquad L_t = \ell_{t_1} + \ldots + \ell_{t_k} + \ell_t ,
\end{equation}
where $\ell_{t_j}$ is the boundary local time acquired between two
successive resettings at $t_{j-1}$ and $t_j$, while $\ell_t$ is the
boundary local time acquired between $t_k$ (the last resetting) and
$t$.  As a consequence, the integral over all intermediate states
should be done with the survival probability
\begin{equation}  \label{eq:exp_qell}
e^{-q L_t} = e^{-q\ell_{t_1}} \ldots e^{-q\ell_{t_k}} \, e^{-q\ell_t} ,
\end{equation}
i.e., each resetting is ``penalized'' by the corresponding factor
$e^{-q\ell_j}$:
\begin{align}   \nonumber
& G_{q,\phi}(\x,t| \x_0) = \Phi(t) G_q(\x,t|\x_0) \\  \nonumber
& + \int\limits_{0}^{t} dt_1 \phi(t_1)  \int\limits_{\Omega} d\x_1 \int\limits_{0}^{\infty}  d\ell_1
 e^{-q\ell_1} P(\x_1,\ell_1,t_1|\x_0) \\
& \times \Phi(t-t_1) \int\limits_{0}^{\infty} d\ell  e^{-q\ell} P(\x,\ell,t-t_1|\x_1) + \ldots
\end{align}
The integrals over boundary local times eliminate these variables and
allow one to replace full propagators by propagators:
\begin{align*} 
& G_{q,\phi}(\x,t| \x_0) = \Phi(t) G_q(\x,t|\x_0) + \int\limits_0^{t}  dt_1  \phi(t_1)  \\
& \times \int\limits_{\Omega} d\x_1 G_q(\x_1,t_1|\x_0) \Phi(t-t_1) G_q(\x,t-t_1|\x_1) + \ldots 
\end{align*}
In turn, the Laplace transform with respect to $t$ transforms time
convolutions into products.  Finally, the integrals over intermediate
positions $\x_1$, $\x_2$, etc. can be calculated by using the spectral
decomposition (\ref{Gspectral_t}) of the propagators and the
orthonormality of Laplacian eigenfunctions.  In analogy to
Eq. (\ref{eq:Phi_phi_general}), we get
\begin{align*} 
 \tilde{G}_{q,\phi}(\x,p|\x_0) & = \int\limits_{\Omega} d\x' \L_p\{ \Phi(t) G_q(\x',t|\x_0)\}  \\
& \times\sum\limits_{k=0}^\infty \frac{u_k^{(q)}(\x) [u_k^{(q)}(\x')]^*}{1 - \tilde{\phi}(p + D\lambda_k^{(q)})} \,.
\end{align*}
However, the main difference with Eq. (\ref{eq:Phi_phi_general}) is
that the factor $\L_p\{ \Phi(t) G_q(\x',t|\x_0)\}$ itself admits the
same spectral decomposition that allows one to compute the integral
over $\x'$:
\begin{align*} 
 \tilde{G}_{q,\phi}(\x,p|\x_0) & = \sum\limits_{k=0}^\infty u_k^{(q)}(\x) [u_k^{(q)}(\x_0)]^* 
\frac{\tilde{\Phi}(p + D\lambda_k^{(q)})}{1 - \tilde{\phi}(p + D\lambda_k^{(q)})} \,.
\end{align*}
Finally, as $\Phi(t)$ is the integral of $\phi(t)$, one has
$\tilde{\Phi}(p) = (1-\tilde{\phi}(p))/p$ so that the last factor is
simply $1/(p + D\lambda_k^{(q)})$.  Inverting this Laplace transform
with respect to $p$, one gets 
\begin{equation}   \label{eq:G_qphi_trivial}
G_{q,\phi}(\x,t|\x_0) = G_q(\x,t|\x_0)
\end{equation}
for any resetting.  In essence, the above formal derivation reflects
the fact that the product of exponential penalizing factors in
Eq. (\ref{eq:exp_qell}) can be reduced to a single factor $e^{-q
L_t}$, where $L_t$ represents the total boundary local time as if
there was no resetting.

We stress that the identity (\ref{eq:G_qphi_trivial}) is the
consequence of the {\it chosen} mechanism of surface reactions.  In
fact, the constant reactivity can be replaced by other surface
reaction mechanisms \cite{Grebenkov20} that is mathematically
equivalent to replacing the exponential law for the random threshold
$\hat{\ell}$ by another law: $\P\{ \hat{\ell} > \ell\} = \Psi(\ell)$.
The overall ``penalizing'' factor Eq. (\ref{eq:exp_qell}) is now
replaced by
\begin{equation}  \label{eq:exp_qell2}
\Psi(\ell_{t_1})\, \ldots \, \Psi(\ell_{t_k}) \, \Psi(\ell_t), 
\end{equation}
which is in general not equal to $\Psi(L_t)$.  As a consequence,
splitting the total boundary local time $L_t$ into ``pieces''
$\ell_{t_1},\ldots,\ell_{t_k}, \ell_t$ by resettings changes the
penalizing factor from $\Psi(L_t)$ to the product in
Eq. (\ref{eq:exp_qell2}) and thus modifies the generalized propagator
$G_{\Psi,\phi}(\x,t|\x_0)$ under resetting so that
\begin{equation}
G_{\Psi,\phi}(\x,t|\x_0) \ne G_{\Psi}(\x,t|\x_0)
\end{equation}
in general.  It is worth noting that the generalized propagator
$G_{\Psi}(\x,t|\x_0)$, given by Eq. (\ref{GPsi}), does not satisfy the
Robin boundary condition (\ref{eq:Gq_Robin}) and thus does not possess
a spectral expansion on common Laplacian eigenfunctions, like
Eq. (\ref{Gspectral_t}) for the conventional propagator
$G_q(\x,t|\x_0)$.  In particular, one cannot evaluate the integrals
over the intermediate positions $\x_k$ in the same way as we did for
the derivation of Eq. (\ref{eq:Phi_phi_general}).  Finding appropriate
tools to compute the generalized propagator $G_{\Psi,\phi}(\x,t|\x_0)$
under resetting presents an interesting perspective for future
research.

\subsubsection*{Probability density of the first-crossing time}

Moreover, even for the constant reactivity, one can imagine other
settings, for which the boundary local time resetting would affect the
conventional propagator.  For instance, resetting can model a renewal
of the reactive state of the target or of the particle, therefore
erasing former history of their interactions.  For instance, in the
context of a resource depletion model introduced in
\cite{Grebenkov22f}, the particle receives a unit of resources at
each encounter with the target, while a threshold $\ell$ characterizes
the amount of initially available resources.  In this setting, the
first-crossing time $\T_\ell$ defined by Eq. (\ref{eq:Tell_def}) is
the first-depletion time, at which the resources on the target are
exhausted.  In turn, the boundary local time resetting can be
considered as a replenishment of resources to the initial level
$\ell$.  The depletion dynamics is therefore characterized by the
probability density $U_\phi(\ell,t|\x_0)$ of the first-crossing time
under resetting.  In contrast to the no-resetting case
(Sec. \ref{sec:encounter}), the density $U_\phi(\ell,t|\x_0)$ does not
follow from the full propagator and related quantities.  For instance,
$\rho_\phi(\ell,t|\x_0)$ determines the probability $\P\{\ell_t <
\ell\}$ that the boundary local time $\ell_t$ {\it at time} $t$ does
not exceed the level $\ell$; however, as $\ell_t$ is not monotonously
increasing due to resettings, this probability says nothing about the
values of $\ell_{t'}$ at earlier times $t'$.  In other words, the
processes $\ell_t$ and $\ell_t^{\rm max} =
\max\limits_{0<t'<t}\{\ell_{t'}\}$ are not identical anymore.  It is 
therefore the probability law for $\ell_t^{\rm max}$ that determines
the first-crossing time $\T_\ell$.  In analogy with
Eq. (\ref{eqmaitresse_Pw_l}), one can write the renewal-type relation
for the joint probability density of $\X_t$ and $\ell_t^{\rm max}$ as
\begin{widetext}
\begin{equation}   \label{eqmaitresse_Pw_l_bis}
 P_\phi^{\rm max}(\x,\ell,t|\x_0) = \Phi(t) P(\x,\ell,t|\x_0) +  \int\limits_{0}^{t} dt_1 \phi(t_1) 
\int\limits_{\Omega} \int\limits_{0}^{\ell} d\x_1  d\ell_1  P(x_1,\ell_1,t_1|\x_0) \, \Phi(t-t_1) P(\x,\ell,t-t_1|\x_1) +  \ldots 
\end{equation}
\end{widetext}
Once this infinite series is computed, the distribution of the
first-crossing time under resetting can be determined via
\begin{equation}
\P\{ T_\ell > t\} = \P\{ \ell_t^{\rm max} < \ell\} = \int\limits_0^\ell d\ell' \int\limits_\Omega d\x \, P_\phi^{\rm max}(\x,\ell',t|\x_0).
\end{equation}
The ``only'' difference with Eq. (\ref{eqmaitresse_Pw_l}) is that the
integrals over $\ell_k$ in Eq. (\ref{eqmaitresse_Pw_l_bis}) have an
upper limit $\ell$ instead of $\infty$, to ensure that intermediate
boundary local times $\ell_k$ do not exceed the threshold $\ell$.
However, this change does not allow one to replace full propagators by
$G_0(\x,t|\x_0)$, while the spectral expansions of the resulting
integrated full propagators are more sophisticated and do not allow to
simply evaluate integrals over intermediate positions $\x_k$.  This
challenging problem remains unsolved, even for the Poissonian
resetting.

\section{Diffusion on an interval}
\label{sec:1D}

In order to illustrate our general results, we consider diffusion on
an interval $(0,b)$ of length $b$ with partially reactive endpoints
(i.e., $\Gamma = \{0,b\}$).  For this basic example, most quantities
of interest can be found explicitly and easily drawn.  Note that this
one-dimensional setting is equivalent to three-dimensional diffusion
between parallel planes separated by distance $b$.  Similar explicit
computations should be feasible for other common models such as, e.g.,
a disk, a cylinder, a sphere, or a spherical target surrounded by a
concentric spherical boundary (see \cite{Grebenkov20b,Grebenkov20c}).
We stress that former works on resetting dealt with unbounded domains
so that ``finite-size effects'' have been ignored.

The conventional propagator $G_{q}(x,t| x_0)$ is well known and was
reported in different textbooks (see, e.g.,
\cite{Carslaw,Thambynayagam}).  In particular, the Laplace transform
of the propagator is given by (see, e.g., \cite{Grebenkov20b}) 
\begin{equation}  \label{eq:1D_Gq}
\tilde{G}_{q}(x,p| x_0) = \frac{1}{\alpha V D} \times \left \{  \begin{array}{cc}
       v_q(x)v_q(b-x_0) & (0 \leqslant x \leqslant x_0), \\
       v_q(x_0)v_q(b-x) & (x_0 \leqslant x \leqslant b),
       \end{array}
\right.
\end{equation}
where $\alpha = \sqrt{p/D}$ and
\begin{align}
v_q(x) & = q \sinh(\alpha x) + \alpha \cosh(\alpha x) ,\\
V & = (\alpha^2 + q^2) \sinh(\alpha b) + 2 \alpha q \cosh(\alpha b) .
\end{align}
The Laplace-transformed survival probability reads then
\begin{align}  \label{eq:1D_Sq}
& \tilde{S}_q(p|x_0) = \frac{1}{p V} \biggl(\alpha^2 \sinh(\alpha b) \\  \nonumber
& + q^2 \bigl(\sinh(\alpha b) - \sinh(\alpha (b-x_0)) - \sinh(\alpha x_0)\bigr) \\  \nonumber
& + \alpha q \bigl(2\cosh(\alpha b) - \cosh(\alpha(b-x_0)) - \cosh(\alpha x_0)\bigr)\biggr).
\end{align}
In the limit $q\to \infty$, one gets
\begin{equation}   \label{eq:1D_Sinf}
\tilde{S}_{\infty}(p| x_0) = \frac{1}{p} \biggl(1 - \frac{\sinh(\alpha x_0) + \sinh(\alpha (b-x_0))}{\sinh(\alpha b)}\biggr)  \,.
\end{equation}

The Laplace-transformed full propagator was found in
\cite{Grebenkov20b}.  When $0 \leqslant x \leqslant x_0 \leqslant b$,
one has
\footnote{
There was a misprint in Eq. (A.7) of Ref. \cite{Grebenkov20b}: the
sign minus in front of the last term should be replaced by the sign
plus, as in our Eq. (\ref{eq:1D_Pfull}).}
%
\begin{widetext}
\begin{align}  \nonumber
\tilde{P}(x,\ell,p| x_0) &= \tilde{G}_{\infty}(x,p| x_0)\delta(\ell) + \frac{e^{-C\ell}}{D \sinh^2(\alpha b)} \Bigg\{
 \Bigg( \sinh(\alpha(b-x_0)) \sinh(\alpha(b-x)) + \sinh(\alpha x_0) \sinh(\alpha x) \Bigg) \cosh(E\ell) \\    \label{eq:1D_Pfull}
& + \Bigg( \sinh(\alpha(b-x_0)) \sinh(\alpha x) + \sinh(\alpha x_0) \sinh(\alpha (b-x)) \Bigg) \sinh(E\ell)  \Bigg\},
\end{align}
\end{widetext}
with $C = \alpha \coth(\alpha b) $ and $E = \frac{\alpha}{\sinh(\alpha
b)}$ (if $ 0 \leqslant x_0 \leqslant x \leqslant b $, one has to
exchange $x_0$ and $x$).
The probability density of $\ell_t$ is given by
\begin{align} \nonumber
\tilde{\rho}(\ell,p| x_0) & = \tilde{S}_{\infty}(p| x_0) \delta(\ell) + \frac{e^{-(C-E)\ell}}{D}  \\  \label{eq:1D_rhop}
& \times \frac{\cosh(\alpha b)-1}{\alpha \sinh(\alpha b)} \, \frac{\sinh(\alpha x_0) + \sinh(\alpha (b-x_0))}{\sinh(\alpha b)} \,.
\end{align}
We emphasize that the behavior of this density is drastically
different from the classical L\'evy's result for diffusion on the
half-line,
\begin{equation}  \label{eq:rho_Levy}
\rho(\ell,t|x_0) = \erf\biggl(\frac{x_0}{\sqrt{4Dt}}\biggr) \delta(\ell) + \frac{\exp\bigl(-\frac{(x_0+\ell)^2}{4Dt}\bigr)}{\sqrt{\pi Dt}} \,,
\end{equation}
where $\erf(z)$ is the error function (see discussion in
\cite{Grebenkov19b}).  Note that Eq. (\ref{eq:rho_Levy}) can be easily
deduced by taking the limit $b\to \infty$ in Eq. (\ref{eq:1D_rhop})
and computing the inverse Laplace transform.  The full propagator on
the half-line is also known explicitly, see \cite{Grebenkov20c}.
The volume-average of Eq. (\ref{eq:1D_rhop}) yields
\begin{align}  \nonumber
\tilde{\rho}(\ell,p| \circ) & = \frac{1}{b} \int\limits_0^b dx_0 \, \tilde{\rho}(\ell,p| x_0) \\  \label{eq:1D_rho_circ}
& = \tilde{S}_\infty(p|\circ) \delta(\ell) + \frac{e^{-(C-E)\ell}}{bD} \, \frac{2(\cosh(\alpha b)-1)^2}{\alpha^2 \sinh^2(\alpha b)}   \,,
\end{align}
where
\begin{equation}  \label{eq:1D_Sinf_circ}
\tilde{S}_\infty(p|\circ) = \frac{1}{p} \biggl(1 - 2 \frac{\cosh(\alpha b)-1}{\alpha b \, \sinh(\alpha b)}\biggr).
\end{equation}
Note that the symmetry (\ref{eq:Psymmetry}) of the full propagator
implies that
\begin{equation}
\tilde{P}(x,\ell,p|\circ) = \frac{1}{b} \int\limits_0^b dx_0 \underbrace{\tilde{P}(x,\ell,p|x_0)}_{=\tilde{P}(x_0,\ell,p|x)} 
= \frac{1}{b} \tilde{\rho}(\ell,p| x),
\end{equation}
i.e., it is given by Eq. (\ref{eq:1D_rhop}).
Using these expressions, one can construct different quantities under
resetting.  For illustrative purposes, we focus on the Poissonian
resetting $\Phi(t) = e^{-\omega t}$.

We set $b = 1$ and $D = 1$ to fix units of length and time.  In
particular, the diffusion time scale can be estimated as $t_{\rm diff}
= b^2/(D\pi^2)\approx 0.1$, where $\pi^2/b^2$ is the smallest
eigenvalues $\lambda_0^{(\infty)}$ for perfectly reactive endpoints.
Moreover, in the case of inert endpoints, one also has
$\lambda_1^{(0)} = \pi^2/b^2$ that controls the asymptotic approach to
the uniform distribution.  In other words, $t_{\rm diff}$ is a typical
time needed for the diffusing particle to explore the confining
domain.  This time scale distinguishes low ($\omega t_{\rm diff} \ll
1$), moderate ($\omega t_{\rm diff} \sim 1$), and high ($\omega t_{\rm
diff} \gg 1$) resetting rates.  Note also that we usually consider
three values of the reactivity parameter: $q = 0$ (inert target), $q =
1$ (moderately reactive target), and $q = \infty$ (perfectly reactive
target).

Figure \ref{fig:simu} shows a random trajectory $X_t$ of the particle
and its boundary local time $\ell_t$ for two types of resetting
concerning either the position, or the boundary local time.

\subsection{Position resetting}

We start by looking at the effect of position resetting on the
propagator $G_{q,\omega}(x,t|x_0)$, which can be obtained from the
inverse Laplace transform of Eq. (\ref{eqfinale_Gw}).  Even though
both $\tilde{G}_q(x,p|x_0)$ and $\tilde{S}_q(p|x_0)$ are known
explicitly, we compute the inverse Laplace transform numerically by
using the Talbot algorithm.  

\begin{figure}[t!]
\begin{center}
\includegraphics[width=0.49\textwidth]{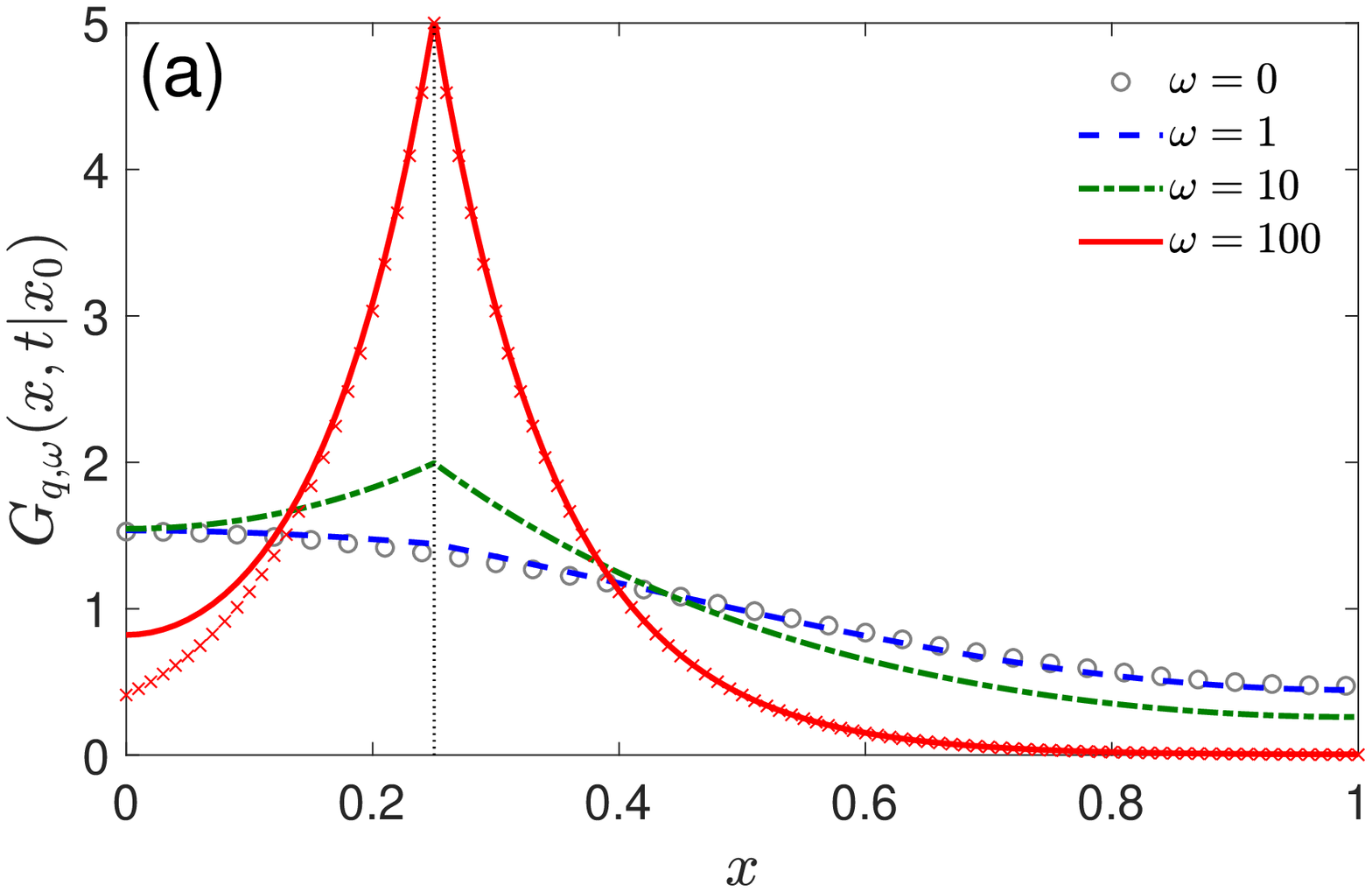} 
\includegraphics[width=0.49\textwidth]{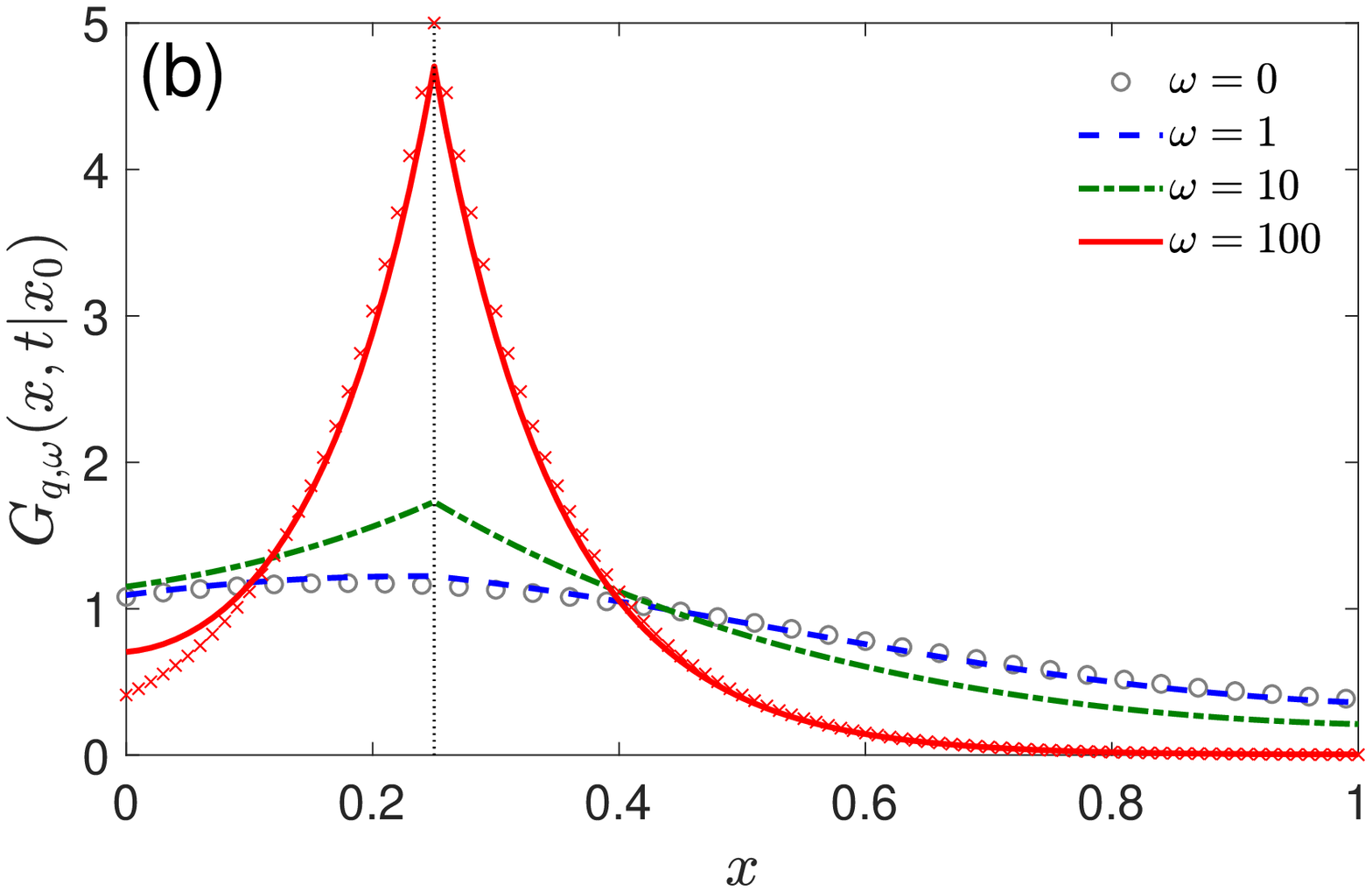} 
\includegraphics[width=0.49\textwidth]{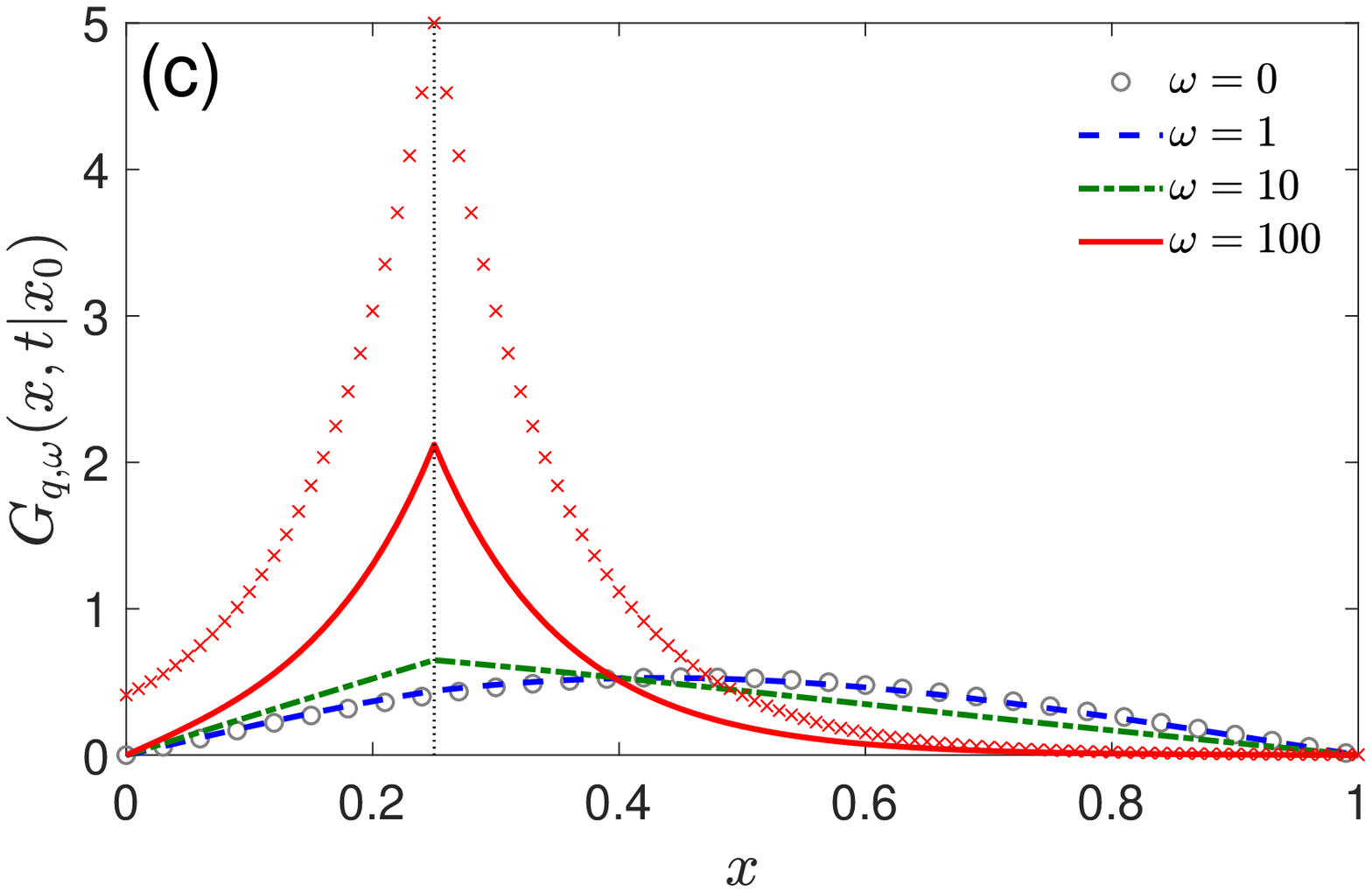} 
\end{center}
\caption{
The propagator $G_{q,\omega}(x,t|x_0)$ under Poissonian resetting of
the position at the rate $\omega$ for diffusion on the unit interval
($b = 1$), with $x_0 = 0.25$, $t = 0.1$, $D = 1$, four values of
$\omega$ (see legend), and $q=0$ {\bf (a)}, $q = 1$ {\bf (b)}, and $q
= \infty$ {\bf (c)}.  Note that $\omega = 0$ corresponds to the
propagator $G_q(x,t|x_0)$ without resetting.  Vertical dotted line
indicates the position $x_0$.  Crosses present the steady-state
density in Eq. (\ref{eq:G_Majumdar}) at $\omega = 100$ for the whole
line without any target.}
\label{Gq_Gqw_q1}
\end{figure}

Figure \ref{Gq_Gqw_q1} illustrates the effect of the resetting rate
$\omega$ onto the propagator $G_{q,\omega}(x,t|x_0)$ at $t = 0.1$
(i.e., at the diffusion time $t_{\rm diff}$) for three values of the
reactivity parameter $q$.  When $\omega = 1$, resettings are too rare
and have little effect on the propagator (we recall that $\omega = 0$
corresponds to the propagator without resetting).  At $\omega = 10$,
there are only few resettings during time $t = 0.1$ but their effect
is clearly seen.  It is further amplified at $\omega = 100$.  When the
target is inert ($q = 0$), resettings prevent the approach of
$G_{0,\omega}(x,t|x_0)$ to the uniform distribution $1$, increasing
the likelihood of finding the particle near its starting point $x_0$
(panel (a)).  In this case, the propagator $G_{0,\omega}(x,t|x_0)$
approaches its steady-state limit $G_\omega^{\rm st}(x|x_0)= \omega
\tilde{G}_0(x,\omega|x_0)$ given by Eq. (\ref{eq:1D_Gq}).  In the
limit $b\to \infty$, this expression tends to the steady-state density
on the positive half-line with reflecting endpoint at the origin:
\begin{equation}
G_\omega^{\rm st}(x|x_0)= \frac{\alpha}{2} e^{-\alpha|x-x_0|} \bigl(1 + e^{-2\alpha\min\{x,x_0\}}\bigr),
\end{equation}
with $\alpha = \sqrt{\omega/D}$.  Moreover, if the reflecting endpoint
is moved to $-\infty$, the last factor approaches $1$, and one
retrieves the seminal result for diffusion on the whole line with
resetting \cite{Evans11}:
\begin{equation}  \label{eq:G_Majumdar}
G_\omega^{\rm st}(x|x_0)= \frac{\alpha}{2} e^{-\alpha|x-x_0|} .
\end{equation}
A distinct cusp-like feature of this function is clearly visible on
Fig. \ref{Gq_Gqw_q1}(a) for $\omega = 10$ and $\omega = 100$.
Moreover, it accurately describes the behavior of the propagator
$G_{0,\omega}(x,t|x_0)$ near $x_0$ at the high resetting rate.

For reactive targets ($q > 0$), the steady-state distribution is zero
in both cases with or without resetting.  In turn, frequent resettings
delay the reaction event and thus prolongate the survival of the
particle.  A cusp-like behavior at high resetting rates $\omega$ is
also present on panels (b) and (c) of Fig. \ref{Gq_Gqw_q1}.

\begin{figure}[t!]
\begin{center}
\includegraphics[width=0.49\textwidth]{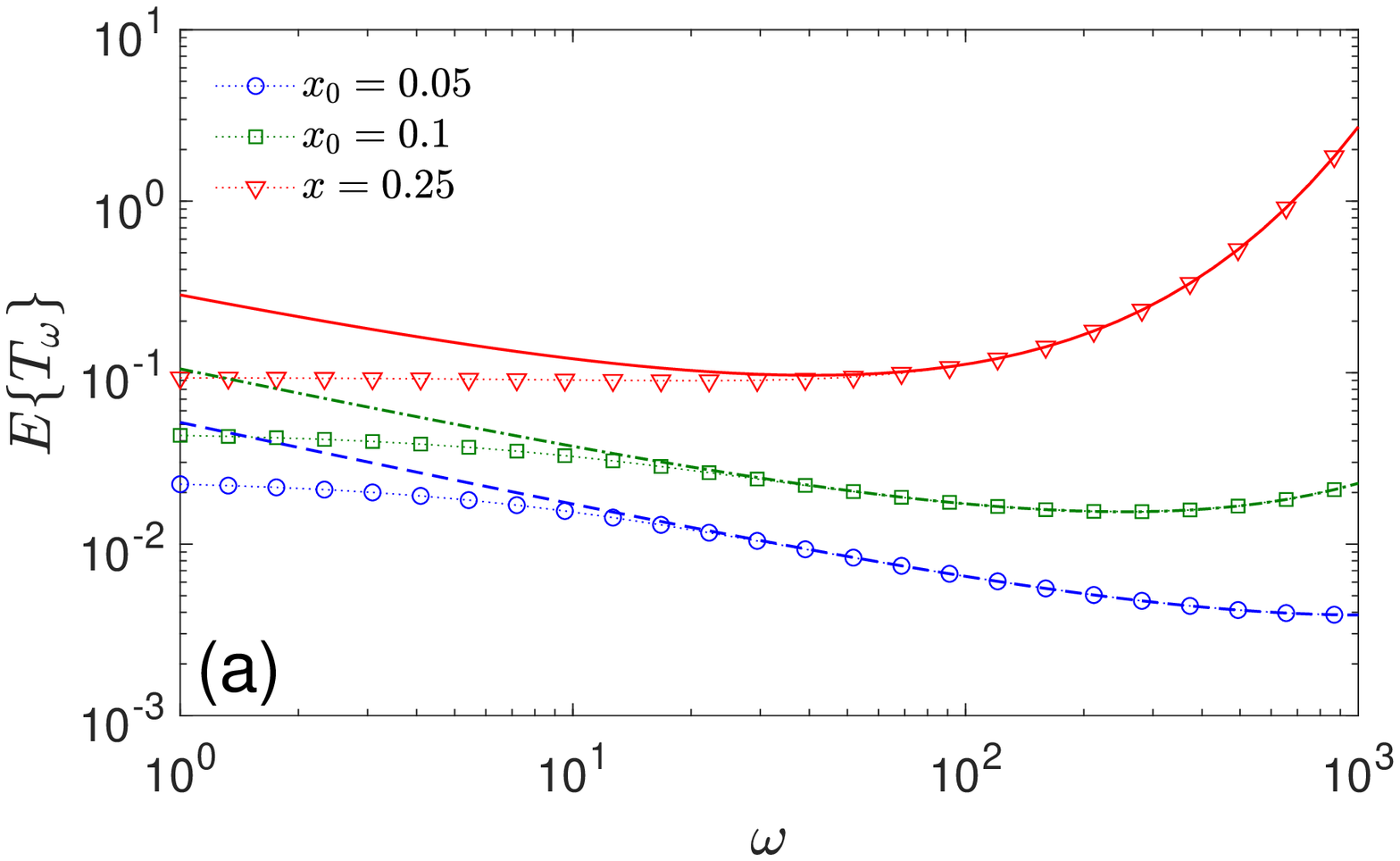} 
\includegraphics[width=0.49\textwidth]{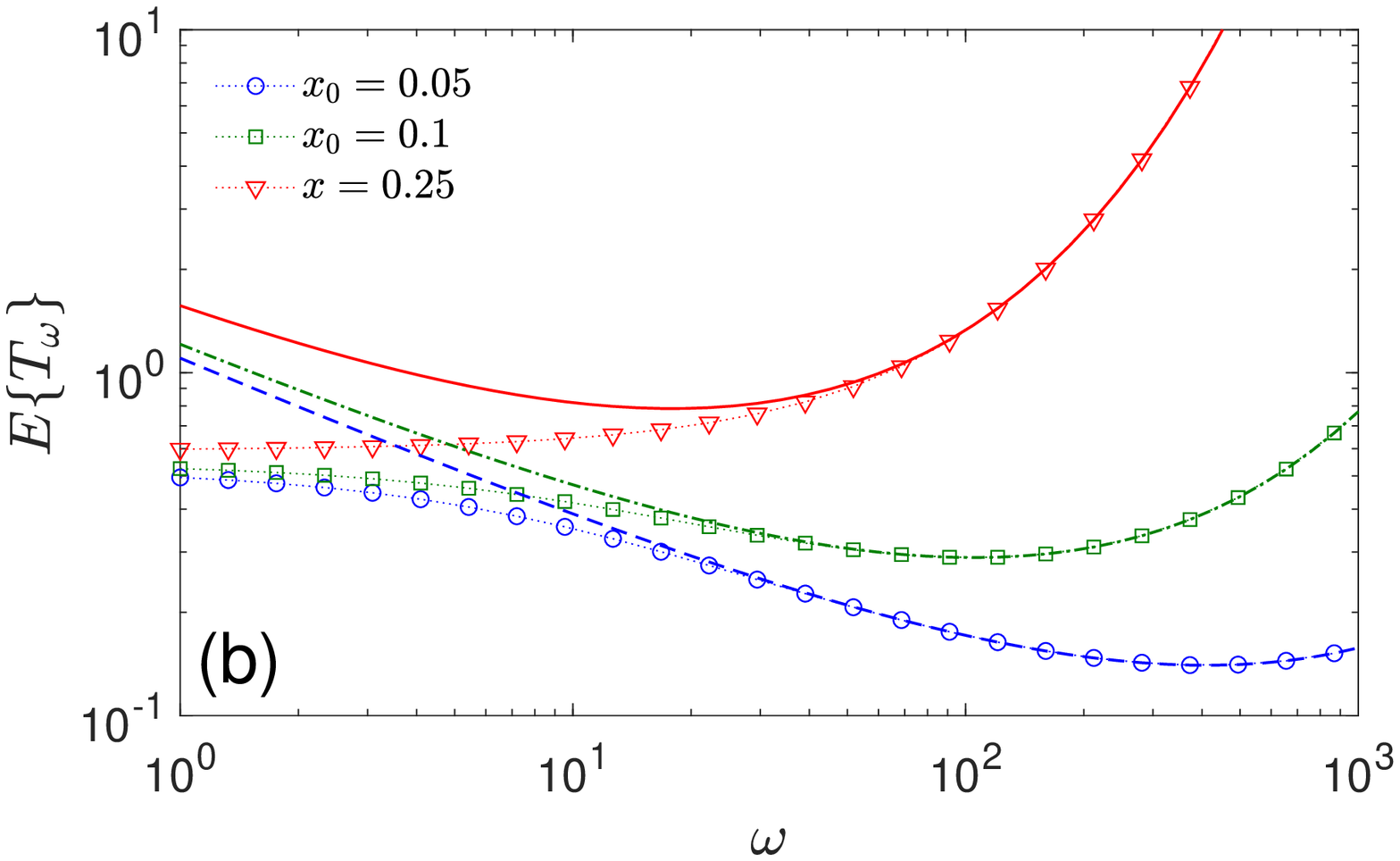} 
\end{center}
\caption{
The mean first-reaction time $\E\{\T_\omega\}$ under Poissonian
resetting of the position at rate $\omega$ for diffusion on the unit
interval ($b = 1$), with $D = 1$, three values of $x_0$ (see the
legend), $q = \infty$ {\bf (a)} and $q = 1$ {\bf (b)}.  Symbols show
the exact result (\ref{eq:Tphi_omega}) with $\tilde{S}_q(p|x_0)$ from
Eq. (\ref{eq:1D_Sq}), while lines illustrate the approximate relation
(\ref{eq:Tomega_1D_app}). }
\label{fig:Tomega}
\end{figure}

This is also consistent with an increase of the mean first-reaction
time $\E\{\T_\omega\}$ as $\omega\to \infty$.  In fact, the
substitution of Eq. (\ref{eq:1D_Sq}) into Eq. (\ref{eq:Tphi_omega})
yields an exact fully explicit expression for $\E\{\T_\omega\}$.  In
the limit $\omega \to 0$, one retrieves the classical result
\begin{equation}  \label{eq:Tomega_0}
\E\{\T_0\} = \frac{x_0(b-x_0)}{2D} + \frac{b}{2qD}
\end{equation}
without resetting.  In turn, if $b\sqrt{\omega/D} \gg 1$, one has 
\begin{equation}
\tilde{S}_q(p|x_0) \approx \frac{1}{p} \biggl(1 - \frac{q}{q+\alpha} e^{-\alpha x_0}\biggr)
\end{equation}
for $0 < x_0 < b/2$ that implies a very simple approximation to the
mean FRT:
\begin{equation}   \label{eq:Tomega_1D_app}
\E\{\T_\omega\} \approx \frac{(1 + \sqrt{\omega/D}/q) e^{x_0 \sqrt{\omega/D}} - 1}{\omega} \,.
\end{equation}
This approximation becomes exact for diffusion on the half-line with
partially reactive origin (i.e., in the limit $b \to \infty$), so that
one retrieves the mean first-passage time reported in \cite{Evans11}
for $q = \infty$ (perfect target) and in \cite{Whitehouse13} for $q <
\infty$.  Note that an additional factor $2$ in front of
$\sqrt{\omega/D}/q$ appears in Eq. (35) from \cite{Whitehouse13} due
to the fact that the target was treated as ``two-sided'', i.e., the
particle diffused on the whole line and could cross the target,
accessing it from the left and from the right; in turn, we consider
that the target is impenetrable and can be accessed only from the
positive semi-axis.

In the limit $\omega \to \infty$, there is an exponentially fast
growth of the mean FRT.  However, if $x_0$ is small enough, this
growth is preceeded by a decay that ensures a minimum of the mean FRT
\cite{Evans11}.  This behavior is very sensitive to the starting
point $x_0$.  Even though the confining domain is bounded here,
resetting to a point near the target can considerably speed up the
search process and the consequent reaction.
Figure \ref{fig:Tomega} illustrates the behavior of the mean FRT for
both perfectly and partially reactive targets.  An approach to a
finite limit (\ref{eq:Tomega_0}) as $\omega\to 0$ is the finite-size
effect due to the boundness of the confining domain, which was not
reported in earlier works.  In particular, if the starting point is
not close to the target (e.g., $x_0 = 0.25$), the minimum of the mean
FRT occurs at $\omega = 0$, i.e., without resetting.  One sees that
finding the optimality range of diffusive search in bounded domains
under resetting can actually be more difficult than in unbounded
domains.

\begin{figure}[t!]
\begin{center}
\includegraphics[width=0.49\textwidth]{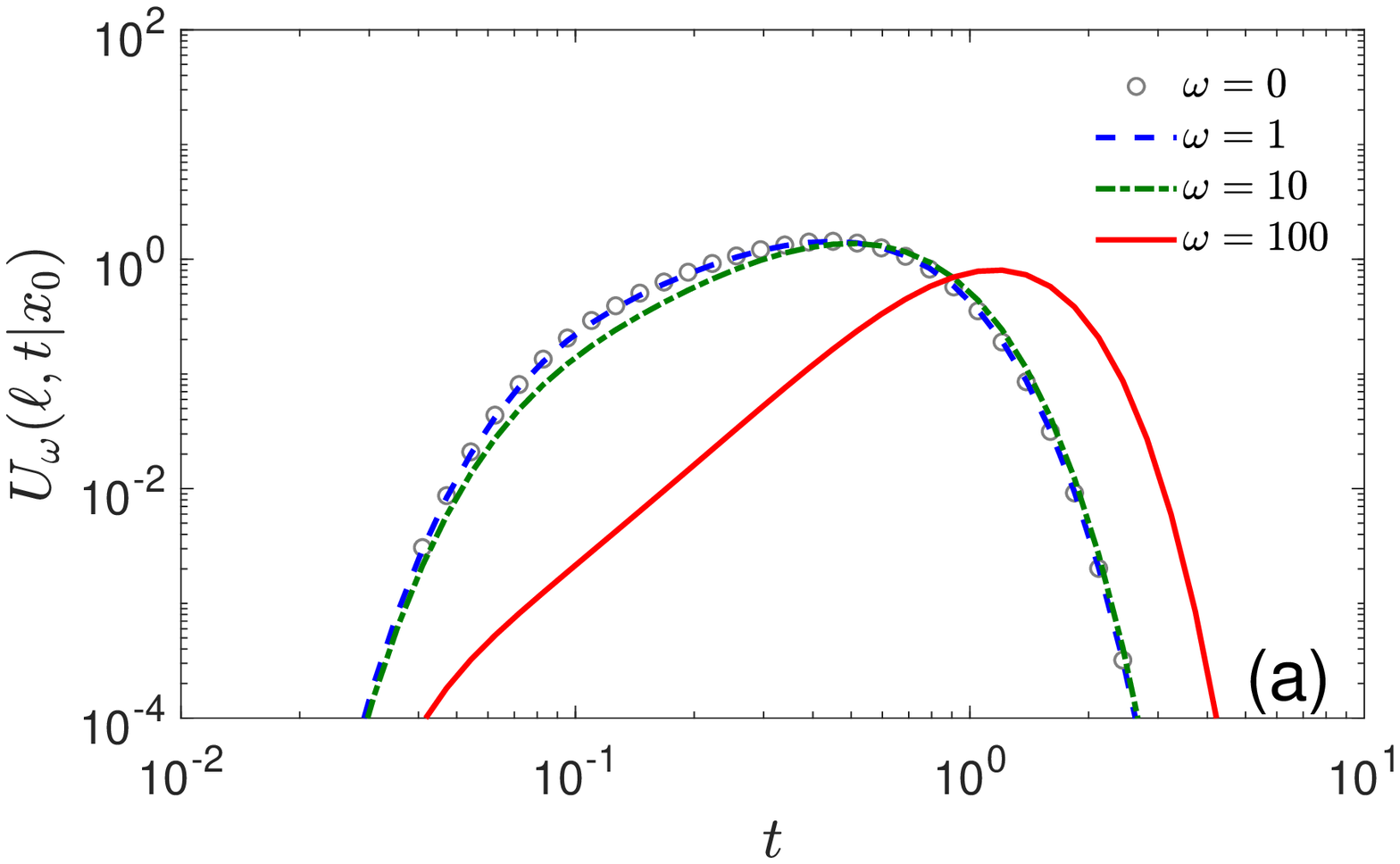} 
\includegraphics[width=0.49\textwidth]{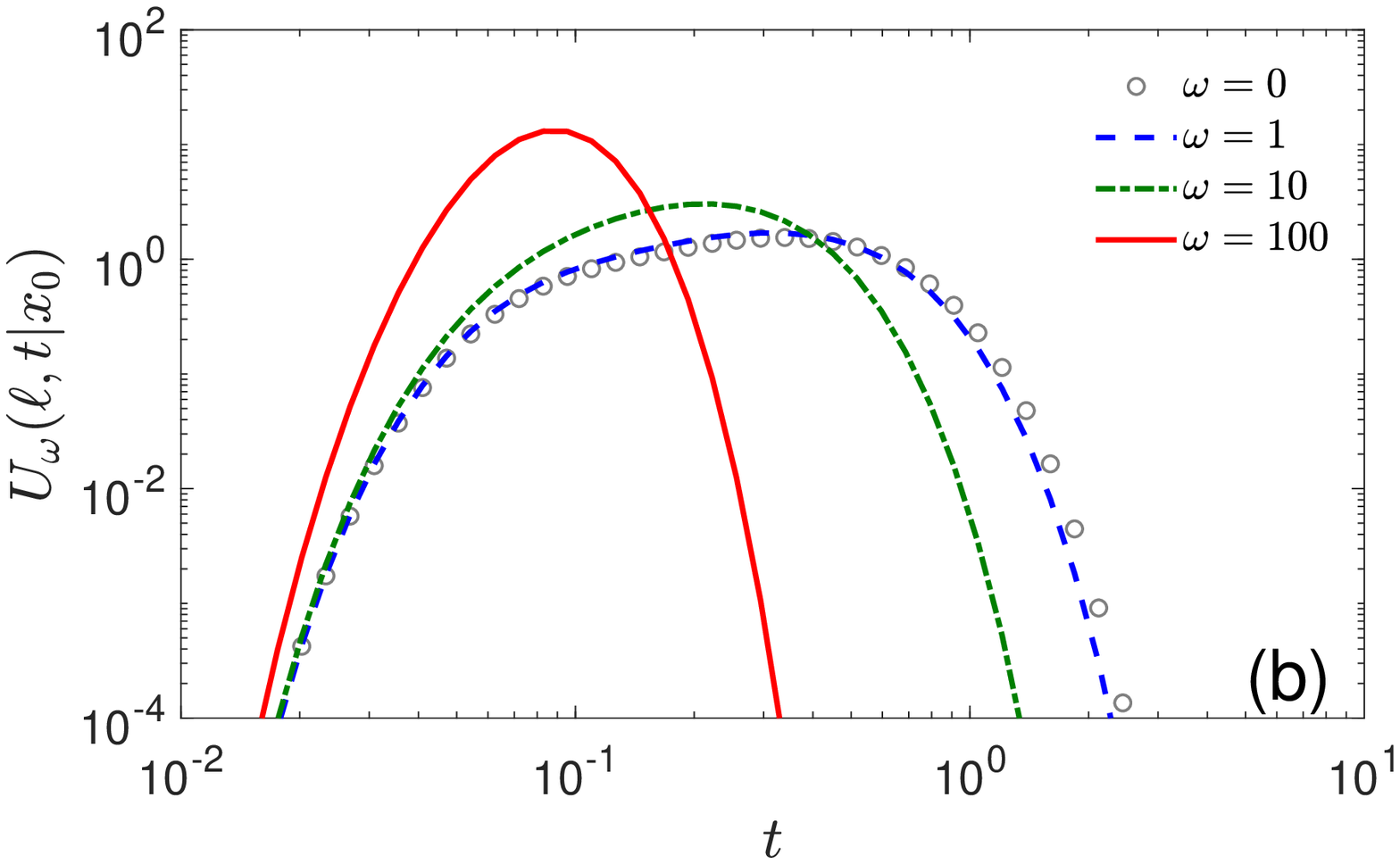} 
\includegraphics[width=0.49\textwidth]{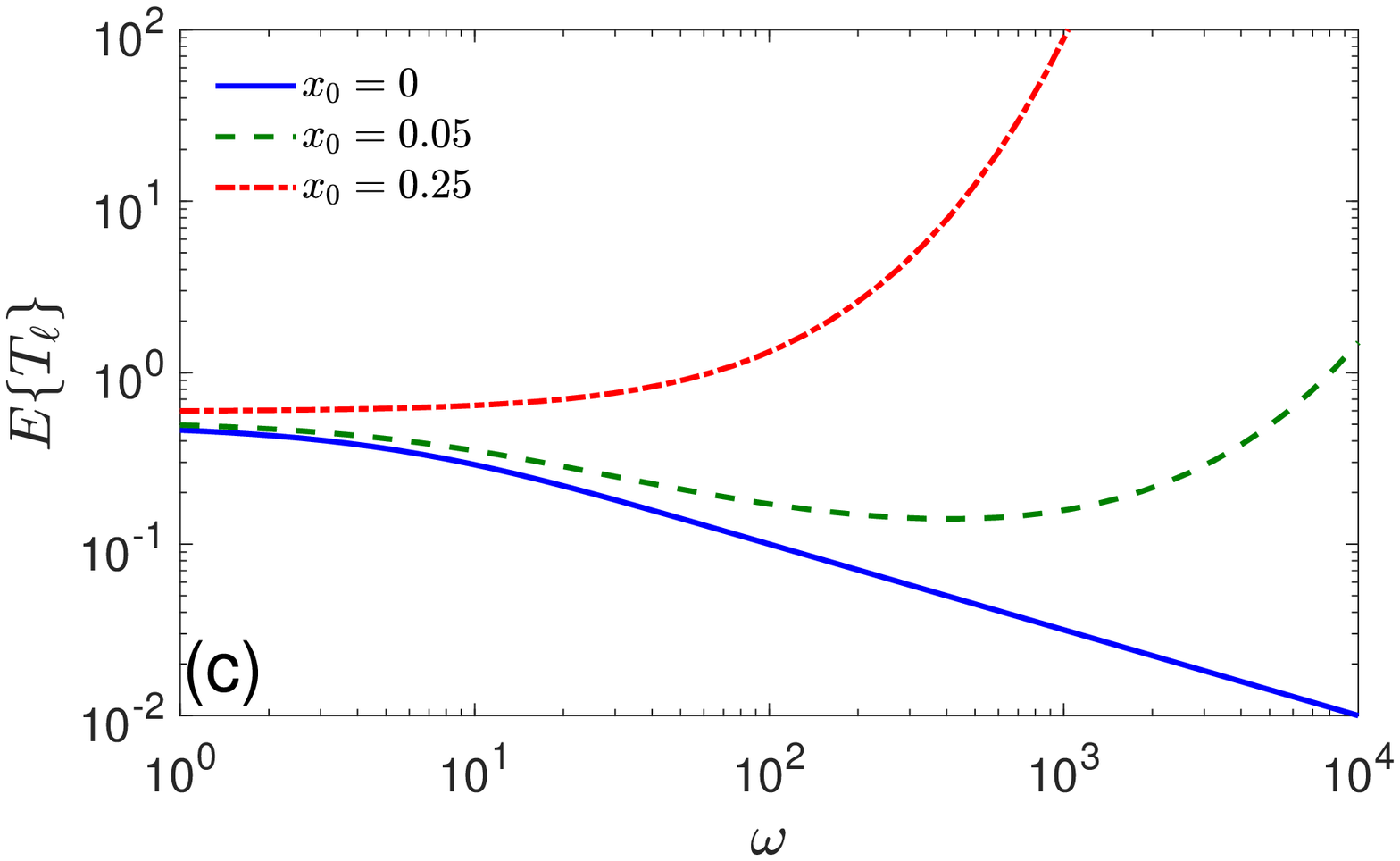} 
\end{center}
\caption{
{\bf (a,b)} The probability density $U_\omega(\ell,t|x_0)$ of the
first-crossing time $\T_\ell$ under Poissonian resetting of the
position at the rate $\omega$ for diffusion on the unit interval ($b =
1$), with $D = 1$, $\ell = 1$, four values of $\omega$ (see the
legend), $x_0 = 0.25$ {\bf (a)} and $x_0 = 0$ {\bf (b)}.  The density
was obtained by the numerical inversion of the Laplace transform in
Eq. (\ref{eq:1D_Uomega}) via the Talbot algorithm.  {\bf (c)} The mean
first-crossing time $\E\{\T_\ell\}$ under Poissonian resetting as a
function of the rate $\omega$, for $\ell = 1$ and three values of
$x_0$ (see the legend).  $\E\{\T_\ell\}$ was obtained as the
derivative of Eq. (\ref{eq:1D_Uomega}) with respect to $p$, evaluated
at $p = 0$.}
\label{fig:Uomega}
\end{figure}

In Appendix \ref{sec:Uomega}, we also computed the Laplace-transformed
probability density $\tilde{U}_\omega(\ell,p|x_0)$ that determines the
first-crossing time under Poissonian resetting of the position.  A
numerical inversion of the Laplace transform in
Eq. (\ref{eq:1D_Uomega}) yields this density in time domain.  Figure
\ref{fig:Uomega}(a,b) illustrates the effect of the Poissonian resetting
onto $U_\omega(\ell,t|x_0)$.  When the starting point $x_0$ is
relatively far from the target ($x_0 = 0.25$, panel (a)), weak
($\omega = 1$) and moderate ($\omega = 10$) resettings have almost no
impact onto the probability density (it remains close to
$U_0(\ell,t|x_0)$ without resetting).  However, at frequent resettings
($\omega = 100$), the particle cannot stay long enough near the target
that slows down the growth of the boundary local time $\ell_t$ and
thus increases the first-crossing time.  The effect of resetting is
opposite when the starting point is located on the target ($x_0 = 0$,
panel (b)).  Here, the particle is reset on the target that speeds up
the growth of the boundary local time and thus decreases the
first-crossing time.  When the starting point is close to the target,
one can therefore expect that resetting can optimize the
first-crossing time $\T_\ell$, in particular, its mean value.  Figure
\ref{fig:Uomega}(c) illustrates this statement.  This behavior will be
studied elsewhere.

\subsection{Boundary local time resetting}

Now we turn the analysis of the Poissonian resetting of the boundary
local time.  The distribution of the boundary local time without
resetting was studied in
\cite{Grebenkov07a,Grebenkov19b,Grebenkov20b,Grebenkov22d} (see also
references therein).  We recall that the probability density
$\rho(\ell,t|\x_0)$ has two contributions: a singular part
$S_\infty(t|\x_0) \delta(\ell)$ from the trajectories that never
reached the target up to time $t$ (and thus $\ell_t = 0$), and a
regular part from the remaining trajectories.  As the first
contribution is trivial, we focus on the regular part.  To avoid
secondary effects of the starting point location, we consider the
volume-averaged probability density $\rho_\omega(\ell,t|\circ)$ that
exhibits similar features.  This quantity is obtained by inversion of
the Laplace transform in Eq. (\ref{eqfinale_rhow_l}) via the Talbot
algorithm, with $\tilde{\rho}(\ell,p|\circ)$ given by
Eq. (\ref{eq:1D_rho_circ}).

\begin{figure}[t!]
\centering
\includegraphics[width=0.45\textwidth]{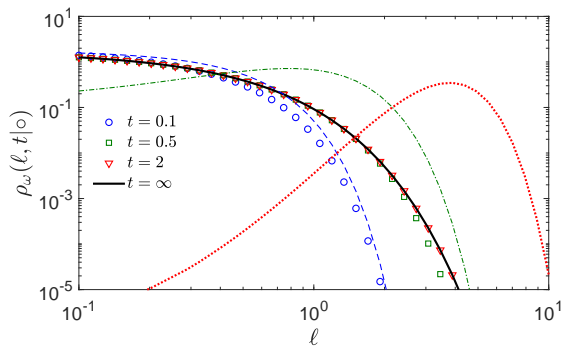} 
\caption{
The regular part of the probability density
$\rho_\omega(\ell,t|\circ)$ of the boundary local time $\ell_t$ under
Poissonian resetting, with $D = 1$, $b = 1$, and several times $t$
(see the legend).  Thin lines show the case without resetting ($\omega =
0$), while symbols present the case with moderate resetting ($\omega =
10$).  A thick black line indicates the steady-state probability
density $\rho_{\omega}^{\rm st}(\ell)$ from
Eq. (\ref{eq:rho_omega_steady}). }
\label{rho_diff_t}
\end{figure}

Figure \ref{rho_diff_t} illustrates the behavior of
$\rho_\omega(\ell,t|\circ)$.  When there is no resetting (thin lines),
the maximum of the probability density $\rho(\ell,t|\circ)$ is
progressively shifted to larger $\ell$.  This is consistent with the
fact that, in a bounded domain, the mean boundary local time grows
linearly with $t$ at long times (see, e.g., \cite{Grebenkov19b}).  In
turn, resetting drastically changes this non-stationary character (see
Sec. \ref{sec:boundary}) so that $\rho_\omega(\ell,t|\circ)$
approaches a steady-state limit $\rho^{\rm st}_\omega(\ell)$ given in
Eq. (\ref{eq:rho_omega_steady}).  This is clearly seen for $t = 0.5$
(squares) and $t = 2$ (triangles), for which
$\rho_\omega(\ell,t|\circ)$ almost coincides with $\rho^{\rm
st}_\omega(\ell)$ (thick black line).  In turn, at moderate time $t =
0.1$, the probability density $\rho_\omega(\ell,t|\circ)$ is still
close to $\rho(\ell,t|\circ)$, i.e., the effect of resetting is weak
at this time scale.

\begin{figure}[t!]
\centering
\includegraphics[width=0.45\textwidth]{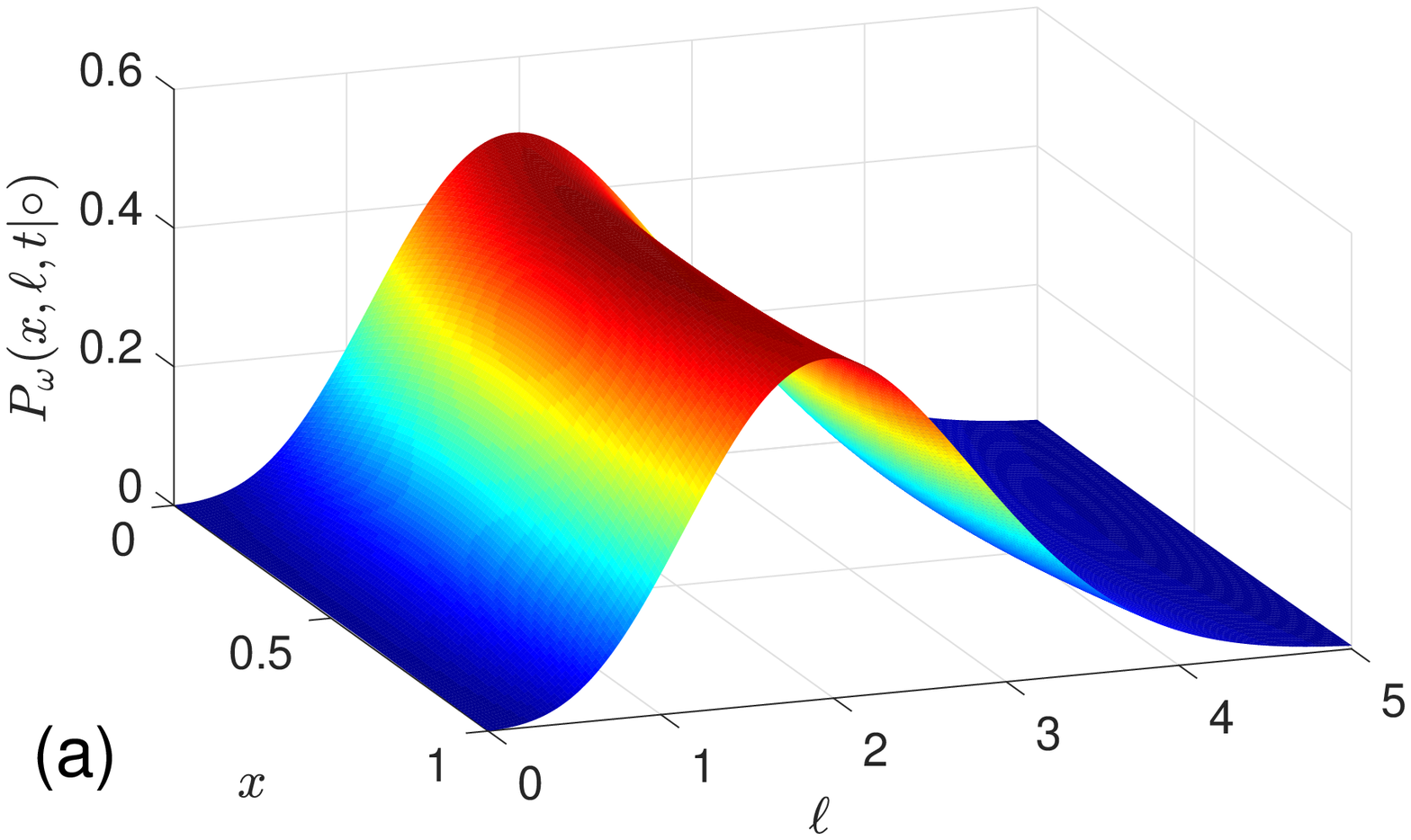} 
\includegraphics[width=0.45\textwidth]{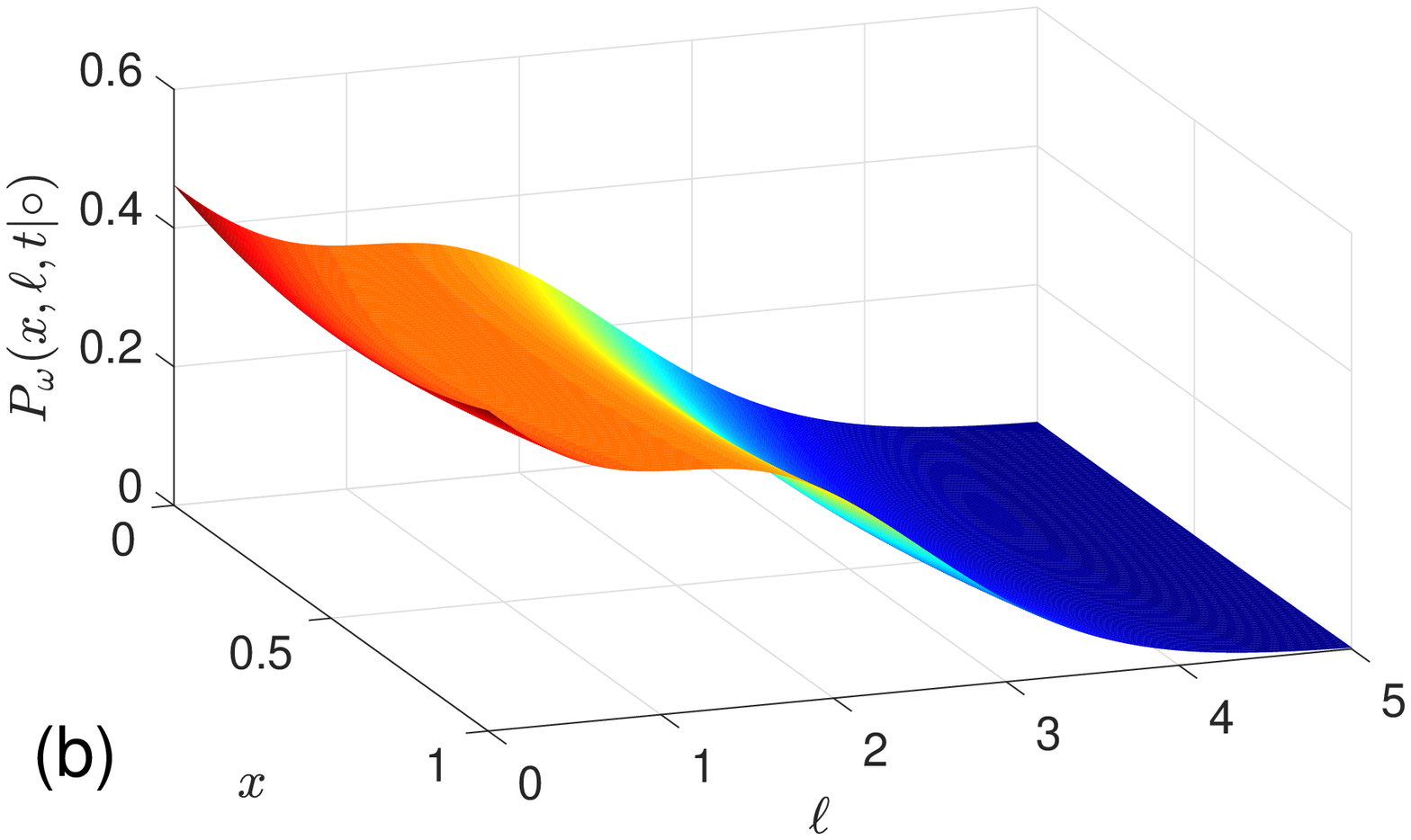} 
\includegraphics[width=0.45\textwidth]{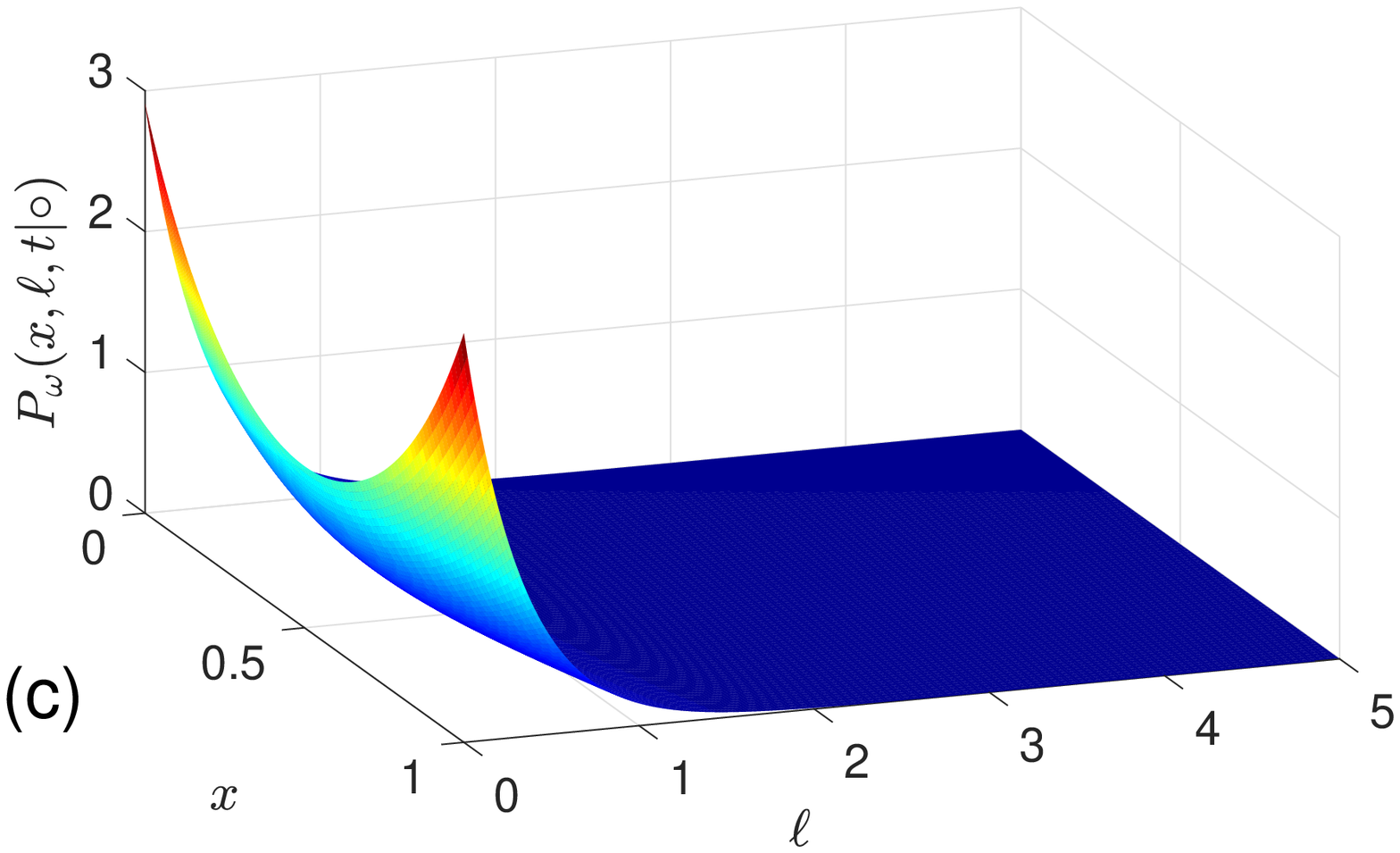} 
\caption{
The regular part of the volume-averaged full propagator
$P_\omega(x,\ell,t|\circ)$ under Poissonian resetting of the boundary
local time at rate $\omega$ for diffusion on the unit interval ($b =
1$), with $D = 1$, $b = 1$, $t = 1$, $\omega = 0$ {\bf (a)}, $\omega =
1$ {\bf (b)}, and $\omega = 10$ {\bf (c)}.  $P_\omega(x,\ell,t|\circ)$
was obtained from Eq. (\ref{eq:tildePomega_circ}) by numerical
inversion of the Laplace transform. }
\label{P_t1}
\end{figure}

\begin{figure}[t!]
\centering
\includegraphics[width=0.49\textwidth]{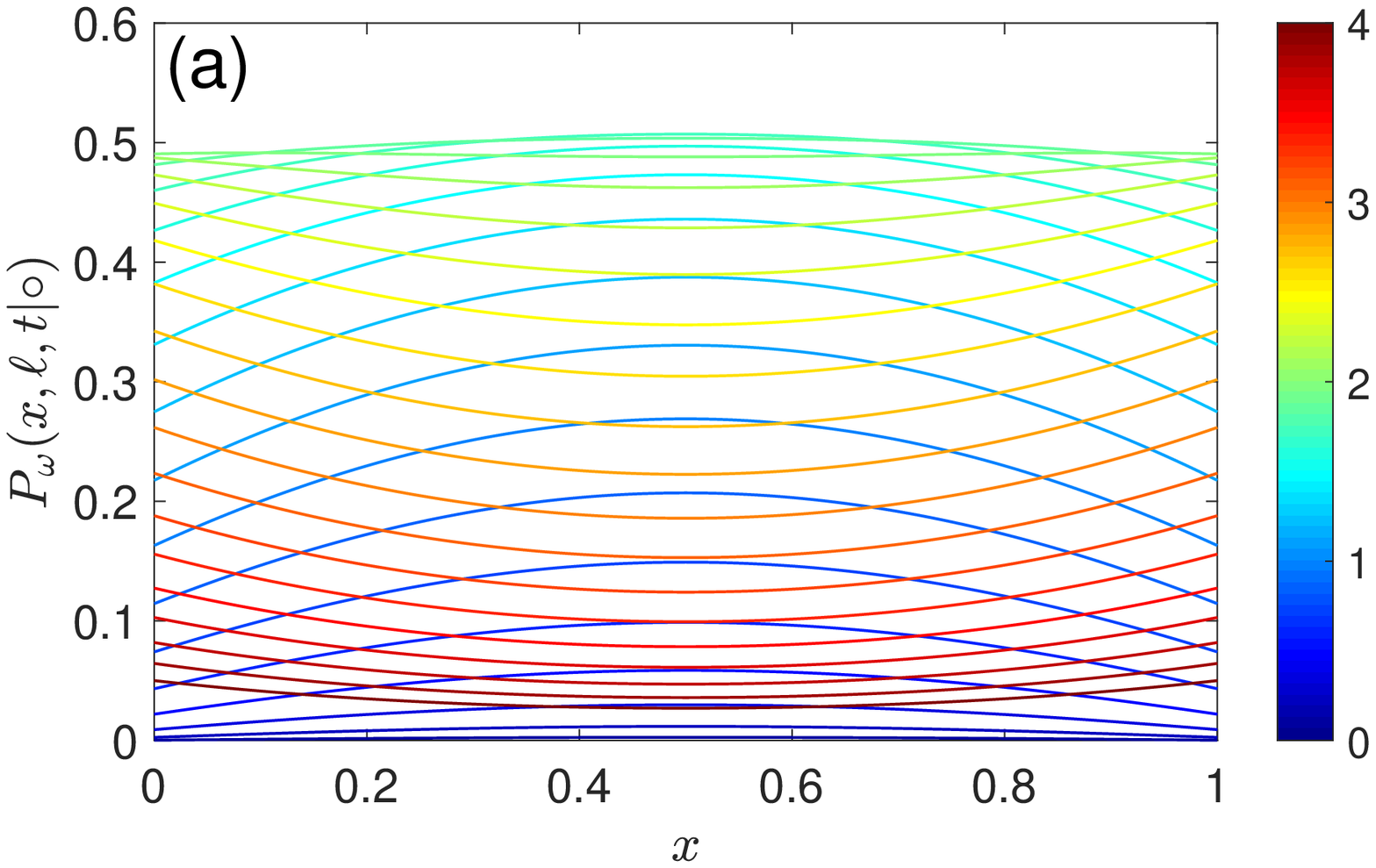} 
\includegraphics[width=0.49\textwidth]{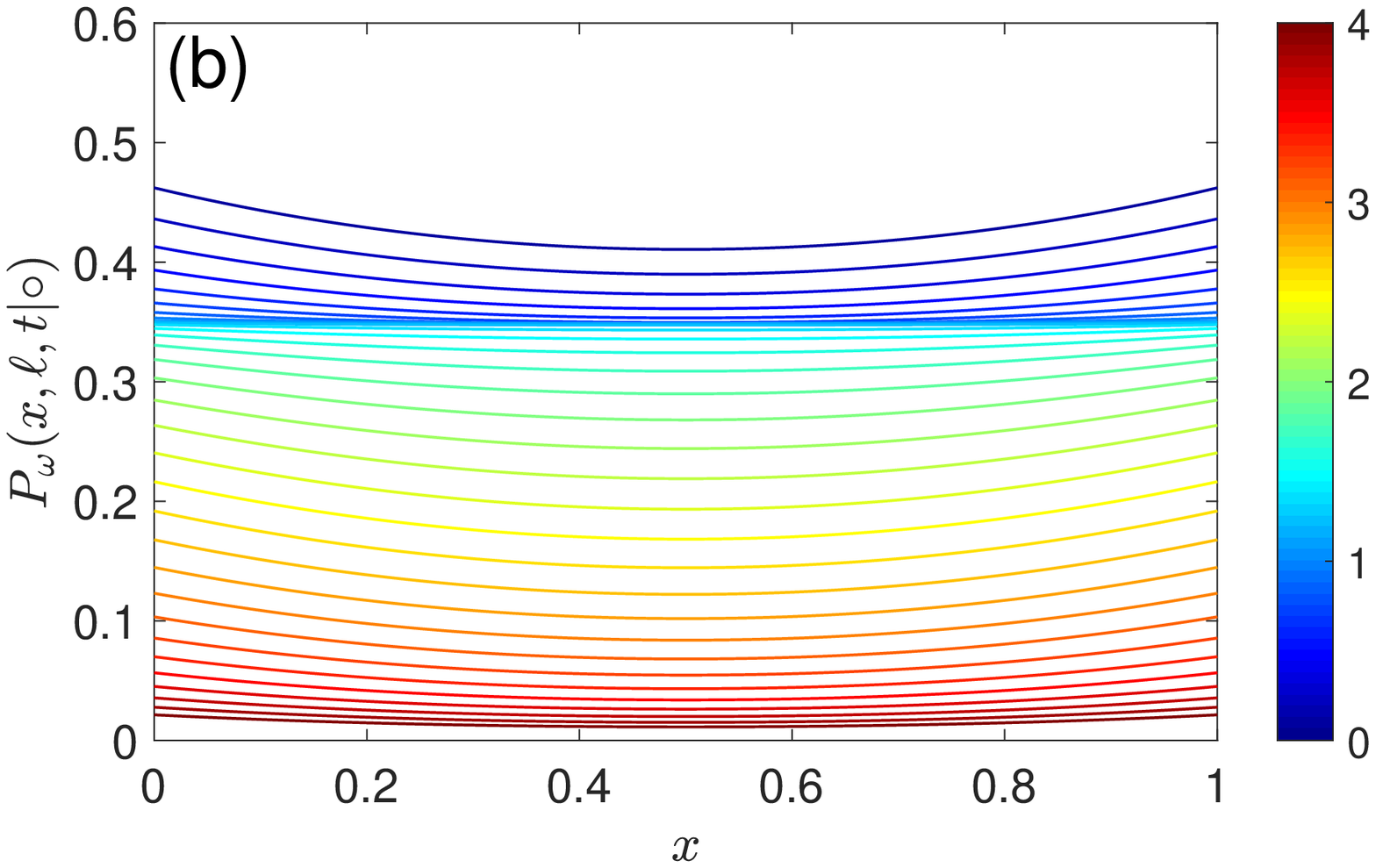} 
\includegraphics[width=0.49\textwidth]{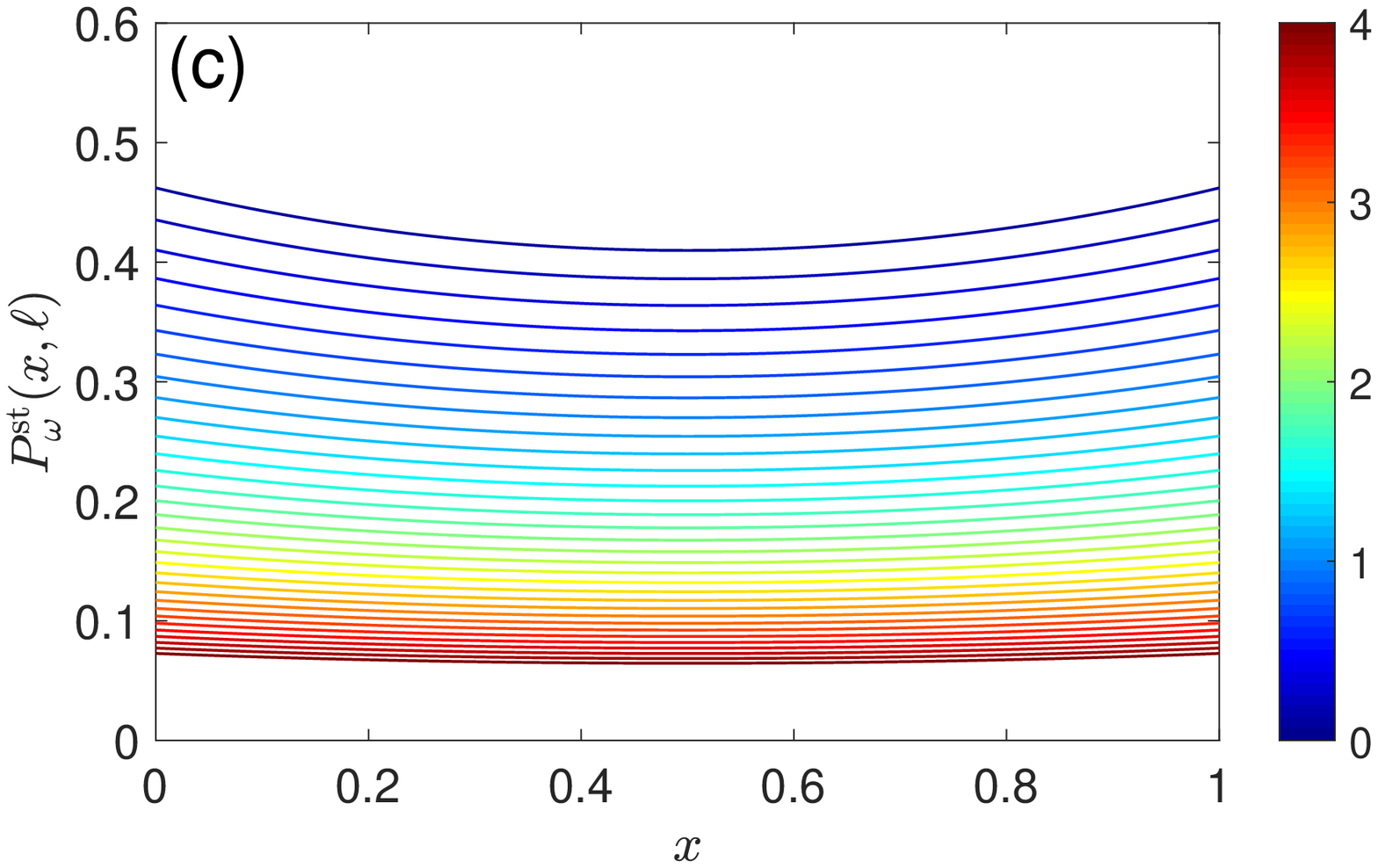} 
\caption{
The regular part of the volume-averaged full propagator
$P_\omega(x,\ell,t|\circ)$ under Poissonian resetting of the boundary
local time at rate $\omega$ for diffusion on the unit interval ($b =
1$).  It is plotted as a function $x$ for multiple values of $\ell$,
with $D = 1$.  Three panels correspond to $t = 1$, $\omega = 0$ {\bf
(a)}, $t = 1$, $\omega = 1$ {\bf (b)}, and $t = \infty$, $\omega = 1$
{\bf (c)}.  33 colored curves correspond to 33 equally spaced values
of $\ell$ ranging from $0$ (dark blue) to $4$ (dark red).}
\label{fig:Px_t1}
\end{figure}

Figure \ref{P_t1} shows the regular part of the volume-averaged full
propagator $P_\omega(x,\ell,t|\circ)$ evaluated at $t = 1$.  In the
case without resetting ($\omega = 0$), this quantity was studied in
\cite{Grebenkov20b}.  Figure \ref{P_t1}(a) shows the expected behavior
of the full propagator without resetting; in particular, one observes
a maximum with respect to the boundary local time $\ell$, which is
progressively shifted to larger $\ell$ as $t$ grows (in other words,
the ``wave'' shown at $t = 1$, moves in the direction of increasing
$\ell$).  As the dependence of $P(x,\ell,t|\circ)$ on the position $x$
is less visible here, we show it more explicitly on
Fig. \ref{fig:Px_t1}(a), where $P(x,\ell,t|\circ)$ is plotted against
$x$ for multiple values of $\ell$ ranging from $0$ to $4$.  As it is
unlikely to get too small values of $\ell$ at $t = 1$,
$P(x,\ell,t|\circ)$ at small $\ell$ (blue curves) has a small
amplitude and exhibits a maximum at the middle of the interval.  In
fact, it is easier for the particle found at the middle to have
smaller values of $\ell$ by encountering the target less frequently.
Similarly, it is unlikely to get too large values of $\ell$ so that
$P(x,\ell,t|\circ)$ at large $\ell$ (red curves) has a small amplitude
and exhibits a minimum at the middle of the interval.  Here, it is
easier for the particle located near the endpoints to encounter the
target more frequently and thus to acquire larger $\ell$.  This
behavior illustrates correlations between $X_t$ and $\ell_t$.

In the presence of resetting, the behavior of the volume-averaged full
propagator $P_\omega(x,\ell,t|\circ)$ is qualitatively different
(Fig. \ref{P_t1}(b,c)).  At large $t$, this propagator reaches the
steady-state distribution $P_\omega^{\rm st}(x,\ell)$, which is given
by Eq. (\ref{eq:Pomega_circ_steady}) and reads for diffusion on the
interval as
\begin{align}
P_{\omega}^{\rm st}(x,\ell) & = \frac{\omega}{b} \biggl(\tilde{S}_\infty(\omega|x) \delta(\ell) +  \frac{\cosh(\alpha b)-1}{\alpha D \sinh^2(\alpha b)} \\
 \nonumber
& \times e^{-\alpha \tanh(\alpha b/2) \ell} \, \bigl(\sinh(\alpha x) + \sinh(\alpha(b-x)) \bigr)\biggr) \,,
\end{align}
with $\alpha = \sqrt{\omega/D}$.  The first (singular) term accounts
for the trajectories that do not reach the target after the last
resetting and thus have $\ell_t = 0$.  In turn, we focus on the second
(regular) term corresponding to $\ell > 0$.  Here, the dependences on
$\ell$ and $x$ are factored out so that $P_{\omega}^{\rm st}(x,\ell)$
always exhibits a minimum at $x = b/2$, whose amplitude is
progressively attenuated as $\ell$ increases.  This behavior is also
illustrated on panels (b) and (c) of Fig. \ref{fig:Px_t1}.

\begin{figure}[t!]
\centering
\includegraphics[width=0.49\textwidth]{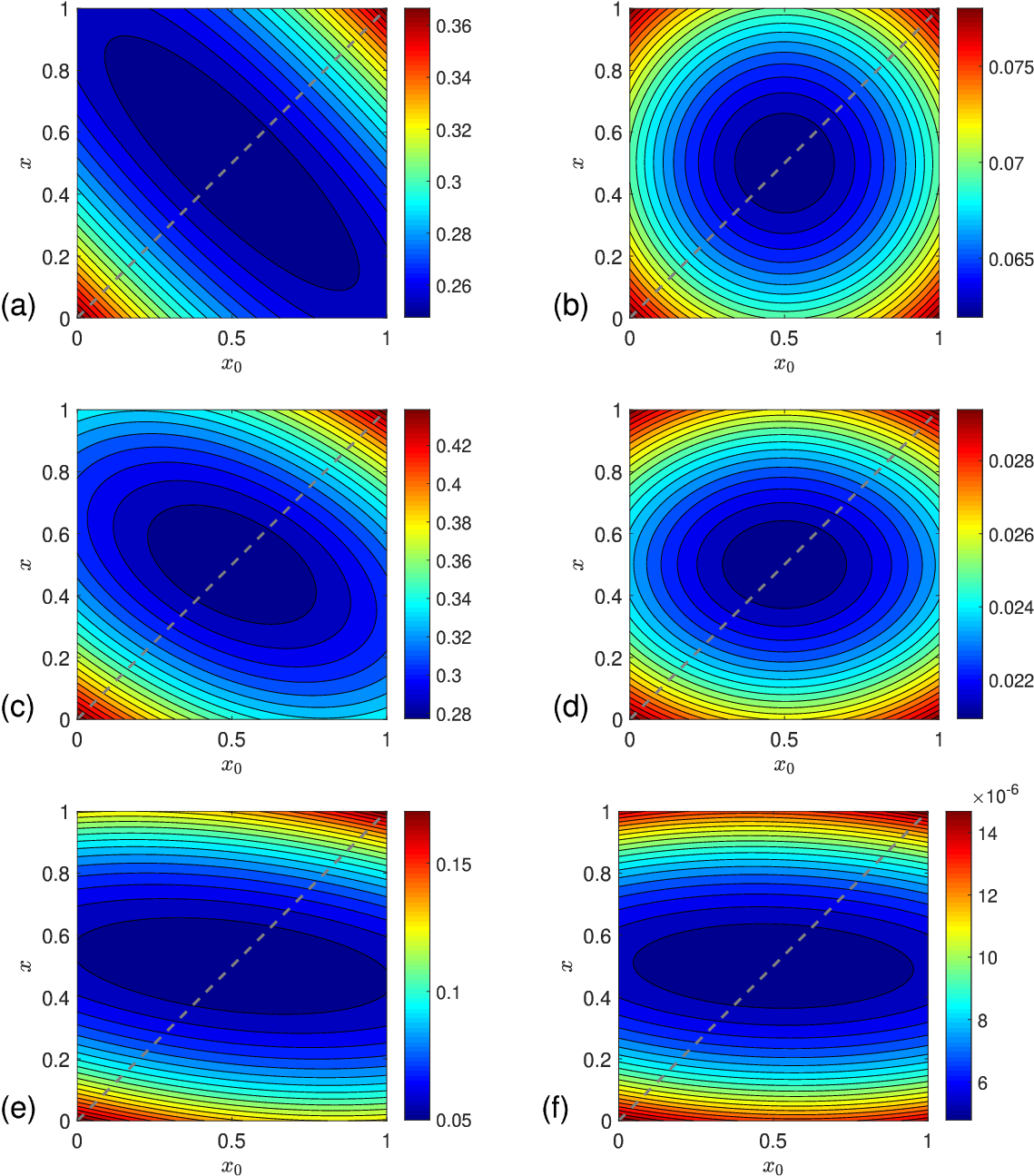}
\caption{
Colored contour plots of the Laplace-transformed full propagator
$\tilde{P}_\omega(x,\ell,p|x_0)$ under Poissonian resetting of the
boundary local time at rate $\omega$ for diffusion on the unit
interval ($b = 1$), with $D = 1$, $p = 1$, $\ell = 1$ (left column)
and $\ell = 4$ (right column).  Panels correspond to $\omega = 0$ {\bf
(a,b)}, $\omega = 1$ {\bf (c,d)}, and $\omega = 10$ {\bf (e,f)}.  }
\label{fig:Pxx0}
\end{figure}

Finally, we illustrate the asymmetry of the full propagator with
respect to the exchange of points $\x_0$ and $\x$ (see
Sec. \ref{sec:boundary}).  Even though the integral in
Eq. (\ref{eqfinale_Pw_l}) can be calculated exactly, the resulting
formulas are too cumbersome so that we perform a numerical
integration.  We also stick to the Laplace domain without performing
its numerical inversion.  Figure \ref{fig:Pxx0} presents a contour
plot of the Laplace-transformed full propagator
$\tilde{P}_\omega(x,\ell,p|x_0)$ at fixed values of $\ell$ and $p$,
and several values of the resetting rate $\omega$.  At $\omega = 0$
(panels (a,b)), this plot is symmetric with respect to the diagonal
(shown by dashed gray line).  In turn, when $\omega > 0$ (panels
(c)-(f)), an asymmetry emerges and is getting more and more visible as
$\omega$ increases.  In particular, the function
$\tilde{P}_\omega(x,\ell,p|x_0)$ at $\omega = 10$ does not almost
depend on $x_0$.

\section{Conclusion}
\label{sec:discussion}

In this paper, we investigated the effect of stochastic resetting onto
diffusion-controlled reactions within the encounter-based approach.
On one hand, this approach disentangles the diffusive dynamics from
surface reactions, yielding a deeper insight onto the search process
and allowing one to implement more sophisticated surface reaction
mechanisms \cite{Grebenkov20}.  On the other hand, it offers richer
opportunities for implementing the stochastic resetting: while former
works were focused on resetting the position of the particle, one can
now investigate the effect of the boundary local time resetting as
well.  We followed these two scenarios and derived two formal
representations (\ref{eq:Pphi_t}, \ref{eq:Phi_phi_general_t}) of the
full propagator $P_\phi(\x,\ell,t|\x_0)$ under arbitrary resetting.
These are new results that allow one to access many other quantities
under resetting such as the conventional propagator
$G_{q,\phi}(\x,t|\x_0)$, the survival probability
$S_{q,\phi}(t|\x_0)$, the probability density $H_{q,\phi}(t|\x_0)$ of
the first-reaction time $\T_\phi$ and its mean value $\E\{\T_\phi\}$,
the probability density $\rho_\phi(\ell,t|\x_0)$ of the boundary local
time $\ell_t$, and the probability density $U_\phi(\ell,t|\x_0)$ of
the first-crossing time $\T_\ell$.  In the case of position resetting,
our general formalism allowed us to deduce some earlier obtained
expressions.  In particular, we retrieved the general expression for
the mean first-reaction time and its asymptotic behavior for
one-dimensional diffusion.  In turn, the scenario of the boundary
local time resetting has been introduced here for the first time so
that all the results that we obtained in Sec. \ref{sec:boundary}, were
not reported earlier.  In particular, we showed that the full
propagator approaches a steady-state distribution and studied its
behavior.  We also observed the asymmetry of the full propagator with
respect to the exchange of the starting and arrival points, which
originates from subtle correlations between the position and the
boundary local time.  Quite counter-intuitively, we found that the
conventional propagator is not affected by any resetting of the
boundary local time.  We showed however that this is a specific
feature of the constant reactivity and the related exponential
distribution in the stopping condition.  This property does not
necessarily hold for other surface reaction mechanisms.  The impact of
stochastic resetting in this more general setting remains to be
uncovered.

While we managed to compute most quantities of interest, at least in
the Laplace domain, the problem of finding the probability density
$U_\phi(\ell,t|\x_0)$ of the first-crossing time under resetting of
the boundary local time remains unsolved.  As described in
Sec. \ref{sec:boundary}, such resetting events destroy the
nondecreasing character of the boundary local time $\ell_t$ so that
$\ell_t$ is no longer equal to its maximum, $\ell_t^{\rm max} =
\max\nolimits_{0<t'<t} \{ \ell_{t'}\}$.  As a consequence, the
knowledge of $\ell_t$ is not enough for describing the probability law
of $\ell_t^{\rm max}$ or, equivalently, the first-crossing time.  Even
though one can still write a renewal-type equation
(\ref{eqmaitresse_Pw_l_bis}) for the joint probability density of
$\X_t$ and $\ell_t^{\rm max}$, we could not find a way to compute the
integrals over intermediate positions $\x_k$ and thus to sum up all
the contributions.  This is a challenging open problem.

Quite naturally, one can go beyond the above two types of resetting.
If both the position and the boundary local time are reset
simultaneously, the full propagator gets a particularly simple form:
\begin{equation}
\tilde{P}_\phi(\x,\ell,p|\x_0) = \frac{\L_p\{ \Phi(t) P(\x,\ell,t|\x_0)\}}{1 - \tilde{\phi}(p)} \,.
\end{equation}
In the case of the Poissonian resetting, one has
\begin{equation}
\tilde{P}_\omega(\x,\ell,p|\x_0) = \biggl(1 + \frac{\omega}{p}\biggr) \tilde{P}(\x,\ell,p+\omega|\x_0) \,,
\end{equation}
which reads in time domain as
\begin{align}  \nonumber
P_\omega(\x,\ell,t|\x_0) & = e^{-\omega t} P(\x,\ell,t|\x_0) \\
& + \int\limits_0^t dt' \, \omega e^{-\omega t'} \, P(\x,\ell,t'|\x_0) \,.
\end{align}
This relation was also derived in \cite{Bressloff22c}, where the case
of simultaneous resettings of the position and the boundary local time
was studied in more details; in particular, the generalized propagator
$\tilde{G}_{\Psi,\omega}(\x,p|\x_0)$ under Poissonian resetting but
arbitrary surface reaction mechanism $\Psi(\ell)$ was obtained.  There
exist however more general forms of asynchronized resettings of $\X_t$
and $\ell_t$, which may considerably affect the full propagator and
the related quantities.  Further explorations in this direction
present an interesting perspective.

As our study was realized independently and in parallel to the work by
Bressloff \cite{Bressloff22c}, it is instructive to highlight several
distinctions between them.  (i) The analysis in \cite{Bressloff22c}
was focused on the Poissonian resetting, which can be implemented by
modifying the partial differential equations for the full propagator
and related quantities; this method presents some advantages, in
particular, Bressloff derived the governing equations for the
propagator $G_{\Psi,\omega}(\x,t|\x_0)$ with a general surface
reaction mechanism; however, its extension to other resetting laws is
difficult.  In turn, we employed the renewal scheme, which is
applicable to any resetting law.  (ii) Bressloff considered two
resetting scenarios: resetting of the position alone and simultaneous
resetting of both $\X_t$ and $\ell_t$; while we also looked at the
first scenario, our main focus was on resetting of the boundary local
time alone, which is technically more difficult.  (iii) The analysis
of \cite{Bressloff22c} was formulated for diffusion outside a compact
obstacle ${\mathcal U}$, i.e., for $\Omega = \R^d \backslash {\mathcal
U}$, in contrast to our focus on bounded domains; even though many
general results are valid in both settings, we often relied on
spectral expansions, which are exclusively applicable in bounded
domains, for which the Laplace operator has a discrete spectrum.  (iv)
Finally, Bressloff considered the exterior of a ball for illustrating
his results, whereas we used a bounded domain (an interval).  We
conclude that these two works provide complementary insights onto the
problem of diffusion-mediated surface phenomena with resetting.

\begin{acknowledgments}
D.S.G. acknowledges the Alexander von Humboldt Foundation for support
within a Bessel Prize award.
\end{acknowledgments}

\appendix

\section{Computation of $U_\omega(\ell,t|x_0)$}
\label{sec:Uomega}

In this Appendix, we detail the computation of the Laplace-transformed
probability density $\tilde{U}_\omega(\ell,p|x_0)$ of the
first-crossing time under Poissonian resetting of the position for
diffusion on the interval.  According to Eq. (\ref{eq:1D_Sq}), we get
\begin{equation}
\tilde{H}_q(p|x_0) = q \frac{q S(x_0) + \alpha C(x_0)}{V} \,,
\end{equation}
where $S(x_0) = \sinh(\alpha(b-x_0))+\sinh(\alpha x_0)$ and $C(x_0) =
\cosh(\alpha(b-x_0))+\cosh(\alpha x_0)$.  We can then express
\begin{align}  \nonumber
\tilde{H}_{q,\omega}(p|x_0) & = \frac{(p+\omega) \tilde{H}_q(p+\omega|x_0)}{p + \omega \tilde{H}_q(p+\omega|x_0)} \\  \label{eq:auxil4}
& = \frac{q (p+\omega) (qS(x_0) + \alpha C(x_0))}{a q^2 + b q + c} \,,
\end{align}
where
\begin{subequations}
\begin{align}
a & = p \sinh(\alpha b) + \omega S(x_0) ,\\
b & = 2p\alpha \cosh(\alpha b) + \omega \alpha C(x_0) , \\
c & = p \alpha^2 \sinh(\alpha b),
\end{align}
\end{subequations}
and we set $\alpha = \sqrt{(p+\omega)/D}$.  Denoting by $q_\pm$ two
roots of the quadratic polynomial in the denominator of
Eq. (\ref{eq:auxil4}),
\begin{equation}
q_\pm = \frac{-b \pm \sqrt{b^2 - 4ac}}{2a} \,,
\end{equation}
one can decompose $\tilde{H}_{q,\omega}(p|x_0)/q$ into a sum of
partial fractions
\begin{align}  \nonumber
\frac{\tilde{H}_{q,\omega}(p|x_0)}{q} & = \frac{(p+\omega)}{a(q_+ - q_-)} 
\biggl(\frac{q_+ S(x_0) + \alpha C(x_0)}{q - q_+} \\
& - \frac{q_- S(x_0) + \alpha C(x_0)}{q - q_-}\biggr) \,.
\end{align}
Substituting this expression into Eq. (\ref{eq:U_phi}), we compute
explicitly the inverse Laplace transform with respect to $q$:
\begin{align} \nonumber
\tilde{U}_\omega(\ell,p|x_0) & = \frac{p+\omega}{a(q_+ - q_-)} \biggl( \bigl(q_+ S(x_0) + \alpha C(x_0)\bigr) e^{q_+ \ell} \\  \label{eq:1D_Uomega}
& - \bigl(q_- S(x_0) + \alpha C(x_0)\bigr) e^{q_- \ell}\biggr) \,.
\end{align}
Since both roots $q_\pm$ are negative, this expression behaves
correctly at large $\ell$.

This is an exact fully explicit expression for the Laplace-transformed
probability density of the first-crossing time under resetting.  Even
though the inverse Laplace transform with respect to $p$ is needed to
get this quantity in time domain, one can investigate the asymptotic
behavior or to compute the moments of the first-crossing time via the
small-$p$ expansion.  For instance, one can check the correct
normalization of the probability density:
$\tilde{U}_\omega(\ell,0|x_0) = 1$.  In the limit $\omega \to 0$, one
has $q_\pm = -\alpha (\cosh(\alpha b) \pm 1)/\sinh(\alpha b)$ that
implies
\begin{equation}
\tilde{U}_0(\ell,p|x_0) = \frac{e^{-\ell \alpha \tanh(\alpha b/2)}}{\sinh(\alpha b)} \bigl(\sinh(\alpha(b-x_0)) + \sinh(\alpha x_0)\bigr) 
\end{equation}
without resetting.

\end{document}